\documentclass[]{emulateapj}

\usepackage{natbib}
\usepackage{bm}
\shorttitle{}
\shortauthors{}

\begin{document}

\title{The Fermi Bubbles: Supersonic AGN Jets with Anisotropic Cosmic Ray Diffusion}

\author{H.-Y.\ K.\ Yang\altaffilmark{1},
M.\ Ruszkowski \altaffilmark{1, 2}, 
P.\ M.\ Ricker\altaffilmark{3},
E.\ Zweibel \altaffilmark{4},
and D.\ Lee \altaffilmark{5}}
\altaffiltext{1}{Department of Astronomy, University of Michigan, Ann Arbor, MI}
\altaffiltext{2}{The Michigan Center for Theoretical Physics}
\altaffiltext{3}{Department of Astronomy, University of Illinois, Urbana, IL}
\altaffiltext{4}{Department of Astronomy and Physics, University of Wisconsin-Madison, Madison, WI}
\altaffiltext{5}{The Flash Center for Computational Science, University of Chicago, Chicago, IL}
\email{Email: hsyang@umich.edu}

\begin{abstract}
  
The {\it Fermi Gamma-ray Space Telescope} reveals two large bubbles in the Galaxy, which extend nearly symmetrically $\sim 50^\circ$ above and below the Galactic center (GC). Using three-dimensional (3D) magnetohydrodynamic (MHD) simulations that self-consistently include the dynamical interaction between cosmic rays (CR) and thermal gas, and anisotropic CR diffusion along the magnetic field lines, we show that the key characteristics of the observed gamma-ray bubbles and the spatially-correlated X-ray features in {\it ROSAT} 1.5\ keV map can be successfully reproduced by a recent jet activity from the central active galactic nucleus (AGN). We find that after taking into account the projection of the 3D bubbles onto the sky, the physical  heights of the bubbles can be much smaller than previously thought, greatly reducing the formation time of the bubbles to about a Myr. 
This relatively small bubble age is needed to reconcile the simulations with the upper limit of bubble ages estimated from the cooling time of high-energy electrons.
No additional physical mechanisms are required to suppress large-scale hydrodynamic instabilities because the evolution time is too short for them to develop. The simulated CR bubbles are edge-brightened, which is consistent with the observed projected flat surface brightness distribution. Furthermore, we demonstrate that the sharp edges of the observed bubbles can be due to anisotropic CR diffusion along magnetic field lines that drape around the bubbles during their supersonic expansion, with suppressed perpendicular diffusion across the bubble surface. Possible causes of the slight bends of the {\it Fermi} bubbles to the west are also discussed.

\end{abstract}

\keywords{}


\section{Introduction}

One of the most important findings from the first two years of {\it Fermi Gamma-ray Space Telescope} observations are two large bubbles extending to $\sim 50^\circ$ above and below the Galactic center (GC), with a width of $\sim 40^\circ$ in longitude \citep{Su10, Dobler10}. The {\it Fermi} bubbles are nearly symmetric about the GC, with only slight bends toward the west (negative longitude). They have approximately flat gamma-ray surface brightness with sharp edges. The bubbles emit at $1\lesssim E_\gamma \lesssim 100$\ GeV, and have a spatially-uniform hard spectrum ($dN_\gamma/dE_\gamma \sim E_\gamma^{-2}$). 
The gamma-ray emission could originate from the decay of neutral pions produced during inelastic collisions between cosmic ray (CR) protons and thermal nuclei \citep[the `hadronic' model;][]{Crocker11}, and/or from inverse Compton (IC) scattering of photons in the interstellar radiation field (ISRF) and the cosmic microwave background (CMB) by CR electrons \citep[the `leptonic' model;][]{Su10, Dobler10}.   
The population of CR electrons is invoked to explain the hard-spectrum microwave synchrotron radiation spatially-correlated with the {\it Fermi} bubbles observed by the Wilkinson Microwave Anisotropy Probe (WMAP), also known as the `WMAP haze' \citep{Finkbeiner04, Dobler08, Dobler12}. The edges of the bubbles also coincide with features in the {\it ROSAT} X-ray maps at 1.5\ keV \citep{Snowden97, Su10}.

As discussed in \cite{Su10}, the unique location and morphology of the {\it Fermi} bubbles suggest that they are created by some large episode of energy injection from the GC, such as a nuclear starburst in the last $\sim 10$\ Myr, or a past accretion event onto the central supermassive black hole (SMBH). The latter scenario, i.e., the formation of CR-filled bubbles by recent jet activity of the active galactic nucleus (AGN), is appealing due to several reasons. Relativistic jets from AGN can accelerate cosmic rays to high energies and thus directly provide the source for the gamma-ray emission. The hard, spatially uniform spectrum of the {\it Fermi} bubbles requires the cosmic rays to be transported from the sites where they are generated to where the bubbles are observed today without significant cooling. 
If the gamma-ray emission is primarily due to IC scattering of the ISRF by CR electrons with energies $10\lesssim E_{\rm cr}\lesssim 100$\ GeV, the IC cooling time of $\sim 100$\ GeV electrons poses a stringent upper limit to the age of the bubbles to be at most a few million years old \citep{Su10}.
Assuming the cosmic rays are produced at the GC, to travel to a distance of several kpc within a few Myr, they must be transported very rapidly, at a speed of $v_{\rm transport} \sim 10^4 (l/10\ {\rm kpc})(t_{\rm age}/1\ {\rm Myr})^{-1}\ {\rm km}\ {\rm s}^{-1}$, which is readily achievable by the fast AGN jets. Furthermore, AGN jets are known to be responsible for radio synchrotron emission from extended extragalactic radio sources \citep[e.g.,][]{Scheuer74, Blandford74} and CR-filled bubbles \citep[e.g.,][]{Laing06} observed in massive galaxies and galaxy clusters \citep{McNamara07}, though their gamma-ray emission cannot be easily detected due to limited sensitivity and resolution. The {\it Fermi} bubbles could possibly be an analogy to such cases, and their proximity may provide a special opportunity to study AGN bubbles in the gamma-ray band.

Forming the {\it Fermi} bubbles in this jet scenario has recently been explored using two-dimensional (2D) simulations by \cite{Guo11a}. They found that AGN jets are able to efficiently carry the cosmic rays to several kpc within $\sim 1-3$\ Myr, and the axial ratios of the bubbles can be reproduced when the density contrast of the jets relative to the surroundings is within the range $0.001 \lesssim \eta \lesssim 0.1$. Despite the general success, there are several discrepancies with the observed bubbles. Their simulated bubbles are subject to large-scale hydrodynamic instabilities that induce ripples on the side of the bubbles, in contrast to the rather smooth surface of the observed bubbles. Also, their simulated bubbles have a uniform CR distribution, which would appear to be limb-darkened if projected onto the sky. In order to ameliorate these problems, they proposed shear viscosity as a possible mechanism to suppress the instabilities as well as to produce an edge-brightened CR distribution that is necessary for a flat projected surface brightness profile \citep{Guo11b}.

They also showed that the sharpness of bubble edges requires suppression of CR diffusion across the bubble surface; otherwise, the bubble edges would be substantially smoothed if the cosmic rays diffuse isotropically with typical coefficients in the Galaxy \citep{Guo11a, Guo11b}. The key to account for the required suppression of CR diffusion may be the effect of {\it anisotropic} diffusion, i.e., diffusion of cosmic rays along magnetic field lines with strongly inhibited cross-field diffusion, expected when the gyro-radius of the cosmic rays is much smaller ($\sim 10^{-9}$\ kpc for a 10\ GeV electron in a 4\ $\mu$G magnetic field) than the mean free path between collisions. If during the expansion of the CR bubbles the ambient gas is compressed, resulting in tangential magnetic field lines near the surface, then the suppression of diffusion across the bubble boundaries could be explained. Using hydrodynamic simulations with specified spatial variation in CR diffusivity, \cite{Guo11a} and \cite{Guo11b} showed that the observed sharp edges can indeed be produced if the diffusion coefficient at the bubble surface is significantly smaller than the canonical values. However, as also stressed by these authors, understanding how this process occurs requires detailed modeling of both the magnetic field and anisotropic CR diffusion along the field lines.

To this end, we perform three-dimensional (3D) magnetohydrodynamic (MHD) simulations of the formation of the {\it Fermi} bubbles by CR jet injections from the central SMBH in the Galaxy. Our simulations {\it self-consistently} include the effects of magnetic field, CR advection, dynamical coupling between cosmic rays and thermal gas, and CR diffusion along the field lines. The main objective of this study is to use 3D numerical simulations involving additional physical mechanisms to investigate whether the jet scenario is able to produce bubbles that are consistent with the observed features, including the shape, the flat surface brightness, and the sharp edges. We will highlight the comparisons with the previous work of \cite{Guo11a} regarding the differences between pure hydrodynamics and MHD, and between 2D and 3D simulations, which have significant impact on the predicted bubble morphology. Moreover, we will show that the interplay between anisotropic CR diffusion and magnetic fields can account for the sharp edges of the observed {\it Fermi} bubbles. The structure of this paper is as follows. In \S~\ref{sec:method} we describe the numerical techniques and initial conditions employed. In \S~\ref{sec:result}, we first present characteristics of our simulated CR bubbles and compare them to the morphology of the observed bubbles, and then study the influence of magnetic fields and anisotropic CR diffusion on the sharpness of bubble edges in \S~\ref{sec:diffusion}. We note the parameter dependencies and available observational constraints in \S~\ref{sec:parameter}, discuss possible causes of the slight bends of the observed bubbles in \S~\ref{sec:bend}, and compare our results with X-ray observations in \S~\ref{sec:xray}. Finally, the summary and implications of our findings are given in \S~\ref{sec:conclusion}.


\section{Methodology}
\label{sec:method}

\subsection{Assumptions and Numerical Techniques}
\label{sec:numerical}

We perform 3D MHD simulations of CR injections from the GC using the adaptive-mesh-refinement (AMR) code {\it} FLASH4 \citep{Flash, Dubey08}. The simulation box is 50\ kpc on a side, refined progressively on the GC located at the center of the simulation domain. The resulting coarsest and finest resolution elements are 0.8 and 0.1 kpc, respectively. The diode boundary condition is used, which is similar to the outflow boundary condition but does not allow matter to flow into the domain. The MHD equations are computed using the directionally unsplit staggered mesh (USM) solver \citep{Lee09, Lee12}. The USM algorithm in FLASH4 is based on a finite-volume, high-order Godunov scheme and is combined with a constrained transport method to ensure the divergence-free condition of magnetic fields. The order of the USM algorithm used in our simulations corresponds to the Piecewise-Parabolic Method (PPM; \cite{Colella84}), which is well suited for capturing shocks and contact discontinuities, as needed in our simulations. In order to incorporate cosmic rays in the simulations, we implemented a new CR module in the USM solver in FLASH4. Here we give a brief summary of the assumptions and equations used to simulate cosmic rays, and present the implementation details and numerical tests of the CR module in Appendix \ref{appendix}.

Though each CR particle travels near the speed of light, it is well known that their collective transport speed in the Galaxy is much smaller, due to scattering by magnetic irregularities. When the scattering is significant, cosmic rays are trapped by the magnetic irregularities that are frozen in the plasma. The cosmic rays effectively move with the thermal gas at the local gas velocity, i.e., are `advected'. Cosmic rays may stream through thermal gas in regions where the magnetic field is locally aligned, with streaming speed limited by the Alfv\'{e}n speed \citep{Skilling71, Kulsrud05}. Since the Alfv\'{e}n speed is much smaller than the local gas speed (as the jet is supersonic), in this study we will neglect the effect of CR streaming. Cosmic rays can also diffuse with respect to the thermal gas as they scatter off magnetic inhomogeneities. Though CR diffusion with a typical diffusivity in the Galaxy is too slow to be the main transport mechanism for the cosmic rays in the {\it Fermi} bubbles \citep{Su10, Guo11a}, it can have non-negligible effects on the CR distribution within and at the edges of the bubbles and thus is included in our simulations.

To simulate collisionless cosmic rays together with collisional plasma, a complete description would require solving for the distribution function of cosmic rays, $f({\bm r}, {\bm p}, t)$. Because $f$ depends on three momentum coordinates as well as position and time, the system of equations is numerically more difficult than a set of fluid equations. In many situations where $f$ is nearly isotropic in momentum space, $f$ can be solved as a function of only ${\bm r}$, $|{\bm p}|$, and $t$ \citep{Skilling75, Miniati01}. In this paper we adopt a more simplistic approach, which is to treat cosmic rays as a second fluid and solve directly for the evolution of CR pressure $p_{cr}$ as a function of ${\bm r}$ and $t$ \citep{Drury81, Jones90, Ryu03, Mathews08, Rasera08}. The cosmic rays are advected with the thermal gas, and in return the gas can react to the gradients of the CR pressure. In this approach, the cosmic rays are treated as a single species without distinction between electrons and protons. We did not model the CR energy spectrum, and neglected the cooling and heating processes of cosmic rays, such as energy losses due to synchrotron and IC emission, and reacceleration in shocks. We will investigate their effects in a future work.

The MHD equations including CR advection, diffusion, and dynamical coupling between the thermal gas and cosmic rays can be written as
\begin{eqnarray}
&& \frac{\partial \rho}{\partial t} + \nabla \cdot (\rho {\bm v}) = 0,\\
&& \frac{\partial \rho {\bm v}}{\partial t} + \nabla \cdot \left( \rho {\bm v}{\bm v}- \frac{{\bm B}{\bm B}}{4\pi} \right) + \nabla p_{\rm tot} = \rho {\bm g},\\
&& \frac{\partial {\bm B}}{\partial t} - \nabla \times ({\bm v} \times {\bm B}) = 0, \label{eq:ind}\\
&& \frac{\partial e}{\partial t} + \nabla \cdot \left[ (e+p_{\rm tot}){\bm v} - \frac{{\bm B}({\bm B} \cdot {\bm v})}{4\pi} \right] \nonumber \\
&& = \rho {\bm v} \cdot {\bm g} + \nabla \cdot ({\bm \kappa} \cdot \nabla e_{\rm cr}),\\
&& \frac{\partial e_{\rm cr}}{\partial t} + \nabla \cdot (e_{\rm cr} {\bm v}) = -p_{\rm cr} \nabla \cdot {\bm v} + \nabla \cdot ({\bm \kappa} \cdot \nabla e_{\rm cr}), \label{eq:ecr}
\end{eqnarray}
where $\rho$ is the gas density, ${\bm v}$ is the velocity, ${\bm B}$ is the magnetic field, ${\bm g}$ is the gravitational field, ${\bm \kappa}$ is the diffusion tensor, $e_{\rm cr}$ is the CR energy density, and $e=0.5\rho v^2 + e_{\rm th} + e_{\rm cr} + B^2/8\pi$ is the total energy density. The total pressure is $p_{\rm tot} = (\gamma -1)e_{\rm th} + (\gamma_{\rm cr} -1) e_{\rm cr} + B^2/8\pi$, where $e_{\rm th}$ is the internal energy density of the gas, $\gamma=5/3$ is the adiabatic index for ideal gas, and $\gamma_{\rm cr}=4/3$ is the effective adiabatic index of cosmic rays in the relativistic regime \citep{Jubelgas08}.   

In the presence of a magnetic field, the diffusion term can be written as
\begin{eqnarray}
\nabla \cdot ({\bm \kappa} \cdot \nabla e_{\rm cr}) = 
\nabla \cdot (\kappa_\parallel \hat{\bm b} \hat{\bm b} \cdot \nabla e_{\rm cr}) + \nonumber \\
\nabla \cdot [\kappa_\perp ({\bm I} - \hat{\bm b}\hat{\bm b}) \cdot \nabla e_{\rm cr}],
\label{eq:anisotropic}
\end{eqnarray}
where $\kappa_\parallel$ is the diffusion coefficient parallel to the magnetic field, $\kappa_\perp$ is the perpendicular diffusion coefficient, and $\hat{\bm b}$ is the unit vector of the magnetic field \citep{Braginskii65}. 

In discussions of Galactic CR propagation, $\kappa_\parallel$ and $\kappa_\perp$ are defined with the respect to the {\it mean} magnetic field, not the exact field (see also \S~\ref{sec:galmodel}). Perpendicular diffusion is assumed to be due to some combination of fieldline wandering due to a random field component and magnetic field structure on scales less than the CR gyroradius, which allows the cosmic rays to migrate from one fieldline to another \citep{Duffy95, Jokipii99, Ensslin03, Hauff10}. In our implementation, we recognize that the magnetic field is probably structured on scales below our numerical resolution, so $\kappa_\parallel$ and $\kappa_\perp$ are necessarily defined with respect to a mean. In the simulations presented here, we assume that the true field is oriented along the mean field, and so we set $\kappa_\perp = 0$. Since in typical cases the perpendicular diffusion coefficient is much smaller than the parallel diffusion coefficient \citep[e.g.,][]{Ensslin03}, our test results show that including a nonzero but small perpendicular diffusion coefficient does not change our main conclusions.    

When magnetic field lines are sufficiently tangled on small scales, CR diffusion can be approximated to be isotropic, for which the diffusion term in the equations can be expressed as $\kappa_{\rm iso} \nabla^2 e_{\rm cr}$. Assuming isotropic diffusion, typical values of the diffusion coefficient in the Galaxy is observationally constrained to be $\kappa_{\rm iso} \sim (3-5)\times 10^{28}\ \rm{cm}^2\ \rm{s}^{-1}$, which may depend on the magnetic field structure and CR energy \citep{Strong07}. In order to compare the effects of isotropic and anisotropic CR diffusion, especially on the sharpness of the {\it Fermi} bubbles, we perform simulations for both cases with varied diffusion coefficients. Since in a tangled field the effective diffusion coefficient ($\kappa_{\rm iso}$) can be suppressed compared to that along the field lines ($\kappa_\parallel$), likely by a factor of $\sim 3$ \citep{Tao95}, we choose to vary the parallel diffusion coefficient in the range of $(0.4-1.2) \times 10^{29}\ \rm{cm}^2\ \rm{s}^{-1}$. The simulation parameters are summarized in Table \ref{tbl:params}.

\begin{table*}[tp]
\caption{Simulation Parameters}
\begin{center}
\begin{tabular}{ccccc}
\hline
\hline
Run & Magnetic Field & $l_{\rm B}$ (kpc) & CR Diffusion & $\kappa_{\rm iso}$ or $\kappa_\parallel$ ($\rm{cm}^2\ \rm{s}^{-1}$) \\ 
\hline
A & None & - & None & - \\
B & Tangled & 9 & None & - \\
C & Tangled & 9 & Isotropic & $4.0\times10^{28}$ \\
D & Tangled & 9 & Anisotropic & $4.0\times10^{28}$ \\
E & Tangled & 9 & Isotropic & $1.2\times10^{29}$ \\
F & Tangled & 9 & Anisotropic & $1.2\times 10^{29}$ \\
G & Tangled & 1 & None & -  \\
H & Tangled & 1 & Isotropic & $4.0\times 10^{28}$ \\
I & Tangled & 1 & Anisotropic & $4.0\times 10^{28}$ \\
J & Tangled & 1 & Isotropic & $1.2\times 10^{29}$ \\
K & Tangled & 1 & Anisotropic & $1.2\times 10^{29}$ \\
\hline
\end{tabular}
\end{center}
\label{tbl:params}
\end{table*}


\subsection{The Galactic Model}
\label{sec:galmodel}

In order to form the {\it Fermi} bubbles, we inject CR jets from the GC into pre-existing gas in the Galactic halo. X-ray observations of emission or absorption lines have provided evidences for the existence of hot gas with temperature $T\sim 10^6$ K extending out to the virial radius of the Milky Way halo with a scale height of a few tens of kpc \citep{Blitz00, McCammon02, Rasmussen03, Fang06, Bregman07, Grcevich09, Miller12}. Following \cite{Guo11a}, we initialize the hot gaseous halo to be in hydrostatic equilibrium within a fixed Galactic potential \citep{Helmi01}, assuming the gas is initially isothermal. In our 3D simulation domain, the Galactic potential can be written as 
\begin{equation}
\Phi = \Phi_{\rm halo} + \Phi_{\rm disk} + \Phi_{\rm bulge},
\end{equation}
where
\begin{equation}
\Phi_{\rm halo} = v^2_{\rm halo} \ln(r^2 + d_{\rm h}^2)
\end{equation}
is the dark logarithmic halo, 
\begin{equation}
\Phi_{\rm disc} = - \frac{GM_{\rm disc}}{\sqrt{x^2 + y^2 + (a+\sqrt{z^2+b^2})^2}}
\end{equation}
is the Miyamoto-Nagai disk, and
\begin{equation}
\Phi_{\rm bulge} = - \frac{GM_{\rm bulge}}{r+d_{\rm b}}
\end{equation}
is the spherical Hernquist stellar bulge. The distance to the GC is $r=\sqrt{x^2+y^2+z^2}$, and the GC is in the middle of the computational domain extending from $-25$\ kpc to $+25$\ kpc in x-, y-, and z-directions. The adopted parameters for the potential are $v_{\rm halo} = 131.5\ {\rm km}\ {\rm s}^{-1}$, $d_{\rm h} = 12$\ kpc, $M_{\rm disc}=10^{11} M_\odot$, $a=6.5$\ kpc, $b=0.26$\ kpc, $M_{\rm bulge} = 3.4\times 10^{10} M_\odot$ and $d_{\rm b} = 0.7$\ kpc. 

The gas density is related to the electron number density by 
\begin{equation}
\rho = \mu_e n_e m_{\mu},
\end{equation}
where $m_{\mu}$ is the atomic mass unit, $\mu_e = 5\mu/(2+\mu)$ is the molecular weight per electron, and $\mu=0.61$ is the molecular weight. Given the electron number density at the origin, $n_{e0}$, and the initial gas temperature, $T_0$, the gas density distribution is derived from hydrostatic equilibrium. We choose $n_{e0} = 2\ \rm{cm}^{-3}$ and $T_0 = 2\times 10^6$ K in order to match the observed hot gas density profile (\cite{Miller12}; see also \S~\ref{sec:parameter}). We neglect rotation in the galactic disk because the rotation period is much longer compared to the estimated age of the {\it Fermi} bubbles. 

The hydrostatic model presented here is not the only possible one. The hot gas distribution in the inner galaxy is also well fit by a galactic wind \citep{Everett08, Everett10}
emanating from a ring with inner galactocentric radius about 3 kpc. Although such a wind would have no effect on the initial expansion of the bubbles, it would provide a
rather different background medium in the later stages. We plan to explore this in future work.

One of the main objectives of this paper is to investigate anisotropic CR diffusion caused by magnetic fields in our Galaxy. To this end we initialized the ambient fields (as opposed to injected magnetic fields from the jets, see \S~\ref{sec:jet}) based on available observational constraints.  
The magnetic field in our Galaxy is composed of a large-scale {\it regular} field (or the {\it mean} field) and a small-scale {\it turbulent} field. The latter is associated with the turbulent interstellar medium, which has a typical coherence length of $\sim 5-50$\ pc (see a recent review by \cite{Noutsos12}). Since this scale is not resolved in our simulation, we will only model explicitly the regular field but not the turbulent component. Therefore, we note that CR diffusion considered in our simulations should be interpreted as an {\it effective} diffusion with respect to the {\it mean} field. 

Observational constraints on the strength and structure of the global Galactic magnetic field (GMF) have been greatly improved over the past decade; however, some details are still unclear. For example, the magnetic fields in the galactic disk is estimated to be a few $\mu$G and enhanced in the spiral arms and near the GC. The direction of the disk field roughly follow directions of the arms, but the number of field reversals is still under debate (see \cite{Brown10} and references therein). The halo field also has a typical magnitude of $\sim 1\ \mu$G. Nevertheless, there has been no agreement on a dipole, quadrupole, or even a large-scale coherent vertical field \citep{Brown10}. Due to the uncertainties in the detailed magnetic-field configurations, instead of imposing a certain shape for the regular field, we model the GMF with a tangled field with random orientations. In this paper we show results from two sets of simulations with different magnetic field coherence lengths, i.e., 9\ kpc (as in the halo) and 1\ kpc (as in the disk), in order to bracket possible cases where the coherence length is larger or smaller than the size of the {\it Fermi} bubbles, respectively. 

The initial tangled field is computed outside the main simulation code and is generated by 3D inverse Fourier transform (FFT) of magnetic field that, in $k$-space, has an amplitude given by \citep{Ruszkowski07}
\begin{equation}
B_k \propto k^{-11/6} \exp(-(k/k_1)^4) \exp(-k_2/k),
\end{equation} 
where $k=(k_x^2+k_y^2+k_z^2)^{1/2}$, and $k_1$ and $k_2$ are the cutoff wavenumbers for large and small $k$, respectively. Given the 3D magnetic power spectrum, ${\mathcal B}(k) \propto |B_k|^2$, the coherence length can be defined as $l_B = 2\pi/k_B$, where the coherent scale in $k$-space is defined as \citep{Ensslin03}
\begin{equation}
k_B \equiv 2 \frac{\int_0^\infty k^2 {\mathcal B}(k) dk}{\int_0^\infty k {\mathcal B}(k) dk}.
\end{equation}
In our simulations with $l_B=9$\ kpc, we set $k_1=0.92$ and $k_2=0.3$; for the case of $l_B=1$\ kpc, $k_1=6.98$ and $k_2=3$ are used. To ensure $\nabla \cdot {\bm B}=0$, the following projection is performed for `cleaning' the field of divergence terms in Fourier space:
\begin{equation}
{\bm B}'({\bm k}) = ({\bm I} - \hat{\bm k} \hat{\bm k}) {\bm B}({\bm k}),
\end{equation}
where ${\bm I}$ is the identity operator. and $\hat{\bm k}$ is the unit vector in $k$-space. We note that this operation does not change the power spectrum of the magnetic field fluctuations.  
After performing the inverse FFT, the magnetic field strength is normalized such that the average field strength is $1\ \mu$G. The initial magnetic field for these two cases in the central region is shown in the left panels of Figure \ref{fig:magfield}.  


\subsection{Jet Injection}
\label{sec:jet}

The feasibility of producing the {\it Fermi} bubbles by AGN jets and parameter variations are studied in detail by \cite{Guo11a}. In order to facilitate comparisons, and in particular to focus on the effects of magnetic fields and anisotropic CR diffusion, we set up our jet model (for thermal gas and CR components) based on their prescriptions but include a few modifications. We also inject magnetic fields along with the jets using the AGN subgrid model of \cite{Sutter12}. Our numerical scheme of jet injection is as follows.  

The bipolar jets are introduced at the GC along the $+z$ and $-z$ direction with a constant velocity $v_{\rm jet}$ for a duration of $t_{\rm jet}$ from the start of the simulation. The direction of the jets are parallel to the rotational axis of the Galaxy, motivated by the symmetry of the observed northern and southern bubbles (except the case in \S~\ref{sec:bend} where we suggest an explanation for the slight bends of the observed bubbles). The thermal gas and CR contents of the jets can be described using six parameters: the thermal gas density $\rho_{\rm j}$, the gas energy density $e_{\rm j}$, the CR energy density $e_{\rm jcr}$, the jet velocity $v_{\rm jet}$, the radius of the jet cross-section $R_{\rm jet}$, and the jet duration $t_{\rm jet}$. The sum of kinetic, thermal, and CR power is thus 
\begin{equation}
P'_{\rm jet} = P_{\rm ke}+P_{\rm th}+P_{\rm cr} =
 \left(\frac{1}{2} \rho_{\rm j}v^2_{\rm jet} + e_{\rm j} + e_{\rm jcr} \right) \pi R^2_{\rm jet} v_{\rm jet}.
\label{eq:power}
\end{equation}
In \cite{Guo11a}, the jets are imposed in a cylinder inside the simulation domain. However, this method would introduce an inconsistency between the distance jets can travel in one timestep and the height of the cylinder, and an artificial redistribution of jet energy inside the cylinder. To avoid these drawbacks, we inject the jets through a disk on the $z=0$ plane with radius $R_{\rm jet}$ using the inflow boundary condition, i.e., updating the ghost cells (in the middle of the computational domain) with the jet parameters at each simulation timestep before solving the MHD equations. In this way variables in the simulation domain are updated self-consistently according to the exact fluxes across the boundary calculated from the USM solver.

Jets from SMBH are generally thought to carry magnetic field, as it plays an important role in jet collimation and acceleration (see \cite{Ferrari98} and references therein), and as suggested by observations of lobes of radio galaxies \citep{Owen00, Kronberg01, Croston05}. AGN are also considered to be a viable agent for distributing the fields into the intergalactic medium (IGM) and intracluster medium (ICM) by relativistic jets \citep[e.g.,][]{Koide99, Contopoulos09, Sutter12}. A realistic representation of the GMF thus requires modeling of the magnetized jets. The details of launching the jets in the relativistic regime is beyond the scope of our study; here we only consider the large-scale effects of the magnetized jets after they have propagated to the resolvable scale. To this end, we inject magnetic fields using the subgrid magnetized jet model of \cite{Sutter12}. This model is based on the magnetic `tower' model of \cite{Li06}, in which the magnetic field is composed of a poloidal flux that presumably comes from the dynamo process in black hole accretion disks, as well as a toroidal component that may be generated by shearing of the poloidal flux lines by differential rotation in the disks. The spatial configuration of the injected field is described in \cite{Sutter12}, to which we refer the readers for more details. During the active phase of the jet, a fraction $f_B$ of $P'_{\rm jet}$ is injected in the form of magnetic energy onto the grid as a source term in the induction equation (right hand side of Eq.\ \ref{eq:ind}). 
The $f_B$ is essentially a free parameter, and therefore in our simulations it is set such that the order of magnitude of the resulting magnetic field in the end of the simulations is consistent with observed values of $\gtrsim 50\ \mu$G at the GC \citep{Crocker10}. We find that this can be achieved using $f_B=10^{-3}$, which means the jets are not magnetically dominated.

A few additional quantities are used to characterize the jets: the density contrast between the thermal gas contained in the jets and the ambient gas, $\eta=\rho_{\rm j}/\rho_{\rm amb}$, the thermal energy contrast $\eta_{\rm e}=e_{\rm j}/e_{\rm amb}$, the total jet power $P_{\rm jet}=(1+f_B)P'_{\rm jet}$, and the total injected energy $E_{\rm jet}=P_{\rm jet} t_{\rm jet}$. Note that since we inject the jets from the GC, the ambient gas density and internal energy are defined at the initial values at the {\it origin}, as opposed to a few kpc away from the GC as in \cite{Guo11a}. Due to the cuspy gas density profiles near the GC (see their Figure 1), our derived $\rho_{\rm amb}$ and $e_{\rm amb}$ are higher than their quoted values. Therefore, the jets in our simulations in general are more powerful than theirs, so as to overcome the ambient gas pressure near the origin. 

We have verified that when using identical implementations and parameters as in \cite{Guo11a}, we are able to reproduce their results. However, as also pointed out by the authors, the observed shape of the {\it Fermi} bubbles can be recovered by multiple parameter combinations due to degeneracies. Moreover, the results (e.g., the derived jet power) are dependent on the initial gas profiles employed. Therefore, in this study we make an effort to incorporate available observational constraints on the jet parameters and initial gas profiles (see \S~\ref{sec:galmodel} and \S~\ref{sec:parameter}). By observing X-ray absorption lines through the hot gaseous halo along many different sight lines in the sky, \cite{Miller12} found that the ratio between OVIII and OVII column densities is enhanced for sight lines that pass through the {\it Fermi} bubbles, indicating gas temperature of a few times $10^7$\ K inside the bubbles, much lower than the temperature produced by the fiducial parameters in \cite{Guo11a}. We find that matching the observed bubble temperature requires slower and wider jets, together with adjusted density contrast and CR energy density in order to maintain the observed shape of the bubbles. To this end, we choose a set of fiducial jet parameters for the simulations presented in this paper. Their values are given in the upper part of Table \ref{tbl:jet params}. Given the fiducial parameters and the initial profile of the Galactic halo, the characteristic quantities of the jets are derived and listed in the bottom part of Table \ref{tbl:jet params}. 
In \S~\ref{sec:parameter} we will discuss the parameter dependence of bubble morphology, and we will show that there are in fact very limited degrees of freedom for the jet parameters given all the available observational constraints on the properties of the {\it Fermi} bubbles.    

\begin{table}[tp]
\caption{Input and Derived Jet Parameters}
\begin{center}
\begin{tabular}{llll}
\hline
\hline
Parameter & Description & Value & Unit \\
\hline
$\eta$ & Density contrast & 0.05 & - \\
$\eta_{\rm e}$ & Energy density contrast & 1.0 & -\\
$e_{\rm jcr}$ & CR energy density & $2.5\times10^{-9}$ & ${\rm erg\ cm}^{-3}$ \\
$v_{\rm jet}$ & Jet speed & 0.025 & $c$ \\
$R_{\rm jet}$ & Radius of cross-section & 0.5 & kpc\\
$t_{\rm jet}$ & Duration of injection & 0.3 & Myr\\
\hline
$n_{\rm ej}$ & Electron number density & 0.1& ${\rm cm}^{-3}$\\
$\rho_{\rm j}$ & Thermal gas densiy & $1.95\times 10^{-25}$ & ${\rm g\ cm}^{-3}$\\
$e_{\rm j}$ & Thermal energy density & $1.59\times 10^{-9}$ & ${\rm erg\ cm}^{-3}$\\
$P_{\rm ke}$ & Kinetic power & $3.08\times 10^{44}$ & ${\rm erg\ s}^{-1}$\\
$P_{\rm th}$ & Thermal power & $8.90\times 10^{42}$ & ${\rm erg\ s}^{-1}$\\
$P_{\rm cr}$ & CR power & $1.40\times 10^{43}$ & ${\rm erg\ s}^{-1}$\\
$P_{\rm B}$ & Magnetic power & $3.31\times 10^{41}$ & ${\rm erg\ s}^{-1}$\\
$P_{\rm jet}$ & Total power & $3.31\times 10^{44}$ & ${\rm erg\ s}^{-1}$\\
$E_{\rm jet}$\footnotemark[1]  & Total injected energy & $3.13\times 10^{57}$ & erg\\
\hline
\multicolumn{4}{l}{\footnotesize \footnotemark[1] The total injected energy by both bipolar jets is $2E_{\rm jet}$.}
\end{tabular}
\end{center}
\label{tbl:jet params}
\end{table}


\section{Results}
\label{sec:result}

\subsection{Morphology of Fermi Bubbles}
\label{sec:morphology}

We first present simulations of the {\it Fermi} bubbles without magnetic field and CR diffusion (i.e., Run $A$), focusing on the comparisons with the previous work of \cite{Guo11a}, namely, the differences between 2D and 3D simulations. In particular, we find that projections of the 3D bubbles are nontrivial and have significant impacts on the interpretations of bubble properties previously derived. 

Firstly, the estimation of the formation time of the {\it Fermi} bubbles are significantly influenced by the consideration of the projection effect. In the simulations, the formation time of the bubbles can be defined as the time when the vertical extent of the simulated bubbles reaches the observed height of $|b|\sim 50^{\circ}$. 
If the bubble sizes are negligible compared to the distance between the Sun and the GC, this condition is equivalent to a vertical bubble size of $\sim 10$\ kpc.
In our simulations, the coordinate transformation from the 3D  simulation domain $(x,y,z)$ to the 2D projected map in Galactic coordinates $(l,b)$ is
\begin{eqnarray}
 \tan l &=& - \frac{x}{y+R_\odot}, \\
 \tan b &=& z \left( \frac{y+R_\odot}{\cos l} \right)^{-1},
 \end{eqnarray}
where $R_\odot=8$\ kpc is the distance from the Sun to the GC, and the bubbles are observed from the Sun's coordinate, defined to be $(x,y,z)=(0,-R_\odot,0)$. If we were to adopt the time when the bubbles reaches $|z|=10$\ kpc as the bubble age, the projected map of the bubbles would have an extent of $|b|\sim 60^\circ - 70^\circ$, because the widths of the bubbles ($\sim 4-5$\ kpc) are actually comparable to $R_\odot$, and therefore the front side (closer to the observer) of the 3D bubbles would be projected to a higher galactic latitude and reach the observed latitude much earlier than what would be naively expected. Taking into account the effect of projection, we find that the condition of $|b|\sim 50^\circ$ corresponds to $|z|\sim 6$\ kpc. 
Therefore, we stop the simulations when the height of the CR bubbles of $e_{\rm cr}\geq 10^{-11}\ {\rm erg\ cm}^{-3}$ along the jet axis reaches 6\ kpc, and define the time $t=t_{\rm Fermi}$ as the current age of the {\it Fermi} bubbles. 
Due to the projection effect combined with the fact that our unprojected bubbles are wider, the age of bubbles we derive is generally shorter than previous estimation by \cite{Guo11a}.

The short age of the bubbles derived has several important implications. Firstly, the {\it Fermi} bubbles are observed in the energy range of $1\lesssim E_\gamma \lesssim 100$\ GeV, which corresponds to relativistic CR electrons with $10 \lesssim E_{\rm cr} \lesssim 100$\ GeV if the gamma-ray emission is produced by IC scattering of the ISRF by these CR electrons \citep{Su10}. Since the IC cooling time of CR electrons at the high-energy end ($\sim 100$\ GeV) is only a few Myr (Figure 28 in \cite{Su10}), this puts an upper limit on the age of the {\it Fermi} bubbles. In our AGN jet-inflated-bubble scenario with the consideration of projection, it only takes 1.2\ Myr for the cosmic rays to travel 6\ kpc from the GC to the observed latitude \citep[note that our jets are slower than those used in][and the bubble formation time would have been much longer if the bubbles had to reach 10\ kpc]{Guo11a}.
Therefore, the constraint from the IC cooling time is naturally satisfied by our `young' bubbles. Later in the section, we will show that the short bubble formation time also has critical influence on the dynamics and morphology of the bubbles as observed today.   

\begin{figure}[tp]
\begin{center}
\includegraphics[scale=1.0]{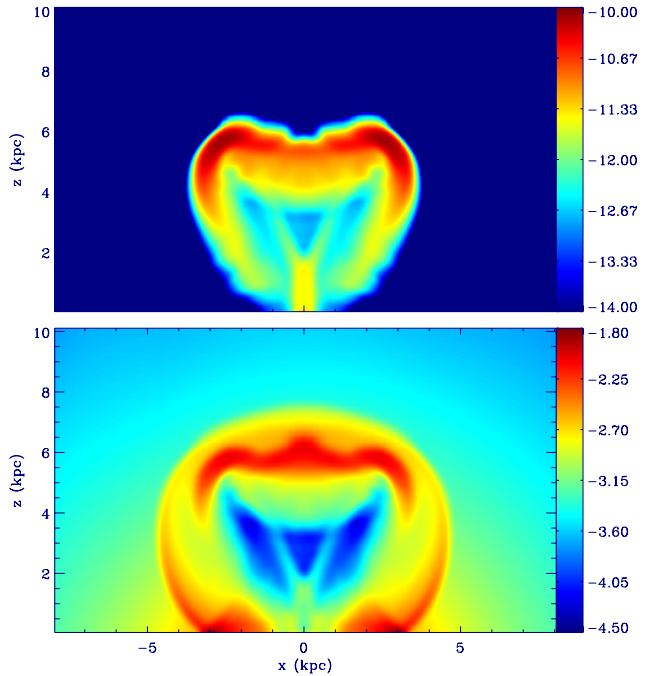} 
\caption{Slices of CR energy density (top) and thermal electron number density (bottom) in simulation coordinates for the fiducial run $A$ at $t_{\rm Fermi}=1.2$\ Myr. Only the northern bubble is plotted since the southern bubble is symmetric with respect to $z=0$. Quantities are shown in logarithmic scale and in {\it cgs} units.} 
\label{fig:fiducial_slice}
\end{center}
\end{figure}

Figure \ref{fig:fiducial_slice} shows the CR energy density and the thermal electron number density sliced through the $y=0$ plane for the fiducial run $A$ at $t_{\rm Fermi}=1.2$\ Myr. In this pure-hydrodynamic run, the bubbles are symmetric with respect to the $z=0$ plane, and thus only the northern bubble is plotted. Initially the jet is over-pressurized with respect to the ambient medium and energetically dominated by the kinetic power. The cosmic rays injected by the central AGN form a pair of CR bubbles above and below the GC. In this run CR diffusion is not included, and thus the evolution of the bubbles is solely due to CR advection and the effects of CR pressure. Due to the high velocity of the thermal gas in the jets, the cosmic rays are rapidly advected to $z=6$\ kpc at the end of the simulation $t_{\rm Fermi}=1.2$\ Myr. The lateral expansion of the bubbles, on the other hand, is mainly driven by the large pressure contrast between the bubbles (contributed by both the thermal gas and the cosmic rays) and the ambient pressure declining with the distance from the GC. Note that the dent on the bubble surface along the jet axis is a result of the cuspy initial gas density profile. Because there is a non-negligible density gradient across the width of the jets, it is more difficult for the central part of the jets to penetrate the dense ambient material at the GC. If a more core-flattened initial density profile were chosen, the top of the bubbles would be flatter and reach 6\ kpc a little earlier, causing a slightly shorter $t_{\rm Fermi}$. However, because the lateral expansion is relatively unaffected, the shape of the projected CR bubbles, which is mainly determined by the side edges closer to the Sun, remains unchanged, and only the projected CR distribution within the bubbles is slightly affected. 
Also in a core-flattened density profile case, the jet would penetrate smaller column density and, thus, smaller jet power would suffice to match observations with simulations.

As can be seen from the bottom panel of Figure \ref{fig:fiducial_slice}, the hot halo gas is expelled and compressed into a shell due to shocks produced by the fast expansion of the CR bubbles. Inside the shocks the electron number density is enhanced to $n_{\rm e}\sim 0.02\ {\rm cm}^{-3}$ and the gas is heated to $T \lesssim 10^8$\ K, separated by a contact discontinuity with the {\it hot, underdense} gas with $n_{\rm e}=10^{-4} \sim 5\times 10^{-3}\ {\rm cm}^{-3}$ and $T \gtrsim 10^7$\ K inside the bubbles. At $t_{\rm Fermi}=1.2$\ Myr, the shocks are moving at a speed of $6000-7000\ {\rm km}\ {\rm s}^{-1}$ (Mach number $M\sim 30$) at the shock front along the $z$ direction and $\sim 2500\ {\rm km}\ {\rm s}^{-1}$ (Mach number $M\sim 12$) in the lateral direction at $z=3$\ kpc. Due to compression and heating, the shocked gas has an enhanced bremsstrahlung emissivity and result in limb-brightened X-ray emission around the bubbles (see \S~\ref{sec:xray}). The strong shocks are in contrast to weak shocks ($M\sim 1-2$) associated with radio bubbles and X-ray cavities in galaxy clusters. Moreover, while in clusters buoyancy is often important in the evolution of radio bubbles, it plays a lesser role for the simulated {\it Fermi }bubbles as their expansion is dominated by the momentum and energy injected by the jets. Note that our prediction of shock velocities at the present day is a few times larger than previous estimation by \cite{Guo11a}. Again because of the short bubble formation time, the shocks are observed today at an earlier epoch of the evolution and have not slowed down as much in the ambient medium. The time for the shock front to move one arcsecond (resolvable with the {\it Chandra} X-ray Observatory) in the lateral direction is $\sim 13$\ yr.   

\begin{figure}[tp]
\begin{center}
\includegraphics[scale=1.0]{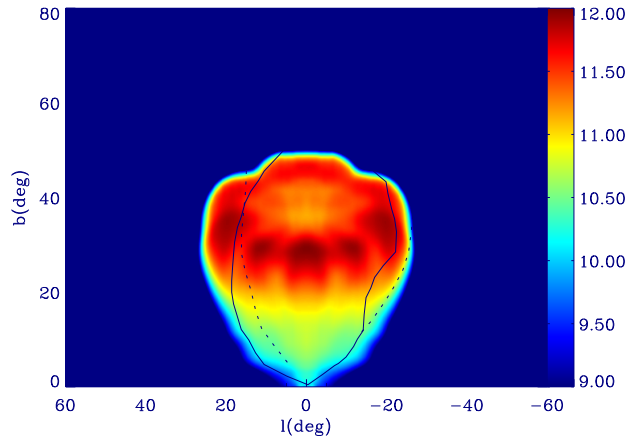} 
\caption{Projected maps of CR energy density in Galactic coordinates for the fiducial run $A$ at $t_{\rm Fermi}=1.2$\ Myr, plotted in logarithmic scale in {\it cgs} units. The solid and dotted lines show the surfaces of the observed northern and southern {\it Fermi} bubbles, respectively.} 
\label{fig:ecrmap}
\end{center}
\end{figure}

The projected maps of CR energy density in Galactic coordinates for the fiducial run $A$ at $t_{\rm Fermi}=1.2$\ Myr are shown in Figure \ref{fig:ecrmap}. The intensity on this map is roughly proportional to the gamma-ray surface brightness assuming the gamma-ray emission is produced by IC scattering by CR electrons of a smoothly distributed ISRF. As can be seen from the plot, the projected CR distribution after evolving for $t_{\rm Fermi}=1.2$\ Myr resembles the shape of the observed {\it Fermi} bubbles (solid and dotted lines for the northern and southern bubbles, respectively; \cite{Su10}), with only slight discrepancies on the east side because of the bends of the observed bubbles (see \S~\ref{sec:bend} for the discussion and modeling of bubble bending). One of the important characteristics of the observed bubbles is their sharp edges. We find that in the absence of CR diffusion (as in Run $A$), the sharpness is also seen for the simulated gamma-ray bubbles, implying suppression of CR diffusion across the bubble surfaces. In \S~\ref{sec:diffusion} we will show that the suppression can possibly be attributed to CR diffusion along magnetic field lines that drape around the bubble surface. 

It is nontrivial to reproduce the flat gamma-ray surface brightness of the observed {\it Fermi} bubbles. 
In order to yield a uniform distribution after line-of-sight projections, the CR energy density must increase toward the bubble boundaries. As shown in the top panel of Figure \ref{fig:fiducial_slice}, our simulated CR energy density indeed is smaller in the bubble interior and gradually increases toward the rims, which then is projected to produce a rather uniform surface brightness (see Figure \ref{fig:ecrmap}), especially along the lateral direction for a fixed Galactic latitude. The simulated CR energy density after projection {\it does} have a dependence on the Galactic latitude $b$ (greater at a higher latitude). Electron aging is an unlikely explanation for the observed flat brightness distribution as a function of latitude, because within the short bubble formation time, 
synchrotron and IC cooling of cosmic rays should be negligible, especially at higher Galactic latitudes (see Figure 28 of \cite{Su10}).
Instead, if the CR electron population is dominant, the theoretical IC radiation distribution, which is proportional to the product of the CR energy density and the ISRF that is stronger near the Galactic plane, may be flatter as a function of $b$ than the emission naively inferred just from the projected CR distribution as a function of $b$. 
Furthermore, the observed surface brightness distribution is subject to larger errors closer to the Galactic disk due to background subtraction. Therefore, the observed flat distribution of gamma-ray emission as a function of latitude is not inconsistent with our simulations. We note that the fiducial parameters adopted in \cite{Guo11a} produce a uniform simulated $e_{\rm cr}$ distribution, which implies a limb-{\it darkened} gamma-ray surface brightness after projection, contrary to the flatness of the observed surface brightness. For this reason, these authors suggested physical viscosity as a possible mechanism to produce a CR distribution consistent with the gamma-ray observation \citep{Guo11b}. Here we show that by using jet parameters appropriate for an observationally-constrained initial gas density profile (see \S~\ref{sec:parameter} for discussion on the chosen parameters), it is possible to reproduce the observed uniform surface brightness without introducing additional physical mechanisms.  

Another feature to note is the rather smooth surface of the observed {\it Fermi} bubbles, which is also reproduced for our fiducial jet parameters. This is in contrast to the results of \cite{Guo11a}, who show that their simulated bubbles are rippled due to large-scale hydrodynamic instabilities. We find that the difference may lie in the estimation of bubble formation time. As discussed in the beginning of this section, our bubble ages are generally shorter than their estimations because it takes much less time for the projected 3D bubbles to match the observed bubble sizes. Given the young age of the bubbles, the hydrodynamic instabilities, including Rayleigh-Taylor (RT) and Kelvin-Helmholtz (KH) instabilities, do not have enough time to develop before the bubbles are observed today (though the instabilities do show if the bubbles are observed at later times). When only the effect of gravity is considered, the timescale for the growth of RT instability at the bubble surface can be estimated by
\begin{equation}
t_{\rm RT} \sim 12.2\ {\rm Myr}\ \sqrt{\frac{1+\eta_{\rm s}}{1-\eta_{\rm s}}} 
\left( \frac{g_{\rm s}}{10^{-8}} \right)^{-1/2} \left(  \frac{\lambda_{\rm s}}{3\ \mbox{kpc}} \right)^{1/2},
\label{eq:RT}
\end{equation}  
where $\eta_{\rm s}$ is the density contrast across the contact discontinuity at the bubble surface, $g_{\rm s}$ is the magnitude of gravitational acceleration in $cgs$ units, and $\lambda_{\rm s}$ is the wavelength of the mode of instability under consideration. For modes comparable to the radius of the bubbles ($\lambda_{\rm s} \sim 3$\ kpc) at the bubble surface on the jet axis ($g_{\rm s} (x=0, y=0, z=6\ {\rm kpc}) \sim 10^{-8}$, $\eta_{\rm s}\sim 0.25$), the RT timescale is $t_{\rm RT}\sim 15.8$\ Myr. This estimate is likely only a lower limit, because the condition for the onset of RT instability becomes more stringent for inflating bubbles \citep{Pizzolato06}. Firstly, the deceleration of bubbles would counteract the gravitational acceleration in the frame of reference comoving with the bubble surface in the early stages of the evolution. Also, the RT growth rate would slow down even further because of an additional drag force as well as an effect of perturbation stretching associated with bubble inflation. Therefore, the RT instability grow on a timescale much longer than the ages of the bubbles, and hence it has a negligible effect.

On the other hand, the KH timescale is more relevant since it is comparable to the estimated ages of the bubbles
\begin{equation}
t_{\rm KH} \sim 1.5\ {\rm Myr} \left( \frac{\lambda_{\rm s}}{3\ {\rm kpc}} \right) 
\left( \frac{\Delta v_{\rm s}}{10^3\ {\rm km\ s}^{-1}} \right)^{-1} 
\left( \frac{\eta_{\rm s}}{0.1} \right)^{-1/2}, 
\end{equation}
where the shear velocity $\Delta v_{\rm s}$ and density contrast $\eta_{\rm s}$ are scaled to typical values at the bubble surface along the lateral direction (e.g., $x=3\ {\rm kpc}, y=0, z=3\ {\rm kpc}$). If $t_{\rm Fermi} > t_{\rm KH}$, the KH instability would grow efficiently and produce ripples in the CR distribution, as found in \cite{Guo11a}; otherwise there is no sufficient time for the instability to develop, and the resulting bubbles should have a smooth surface, as seen in our simulations. Since the ripples are in apparent contradiction to the smooth surface of the observed bubbles, these authors propose to add viscosity as a means to suppress the instabilities \citep{Guo11b}. By taking into account the effect of 3D projection, we show that the formation time of the {\it Fermi} bubbles are in fact shorter than previously expected, which can naturally explain the absence of hydrodynamic instabilities and the smooth surface of the observed bubbles, even when magnetic fields are neglected.  


\begin{figure*}[tp]
\begin{center}
\includegraphics[scale=1.0]{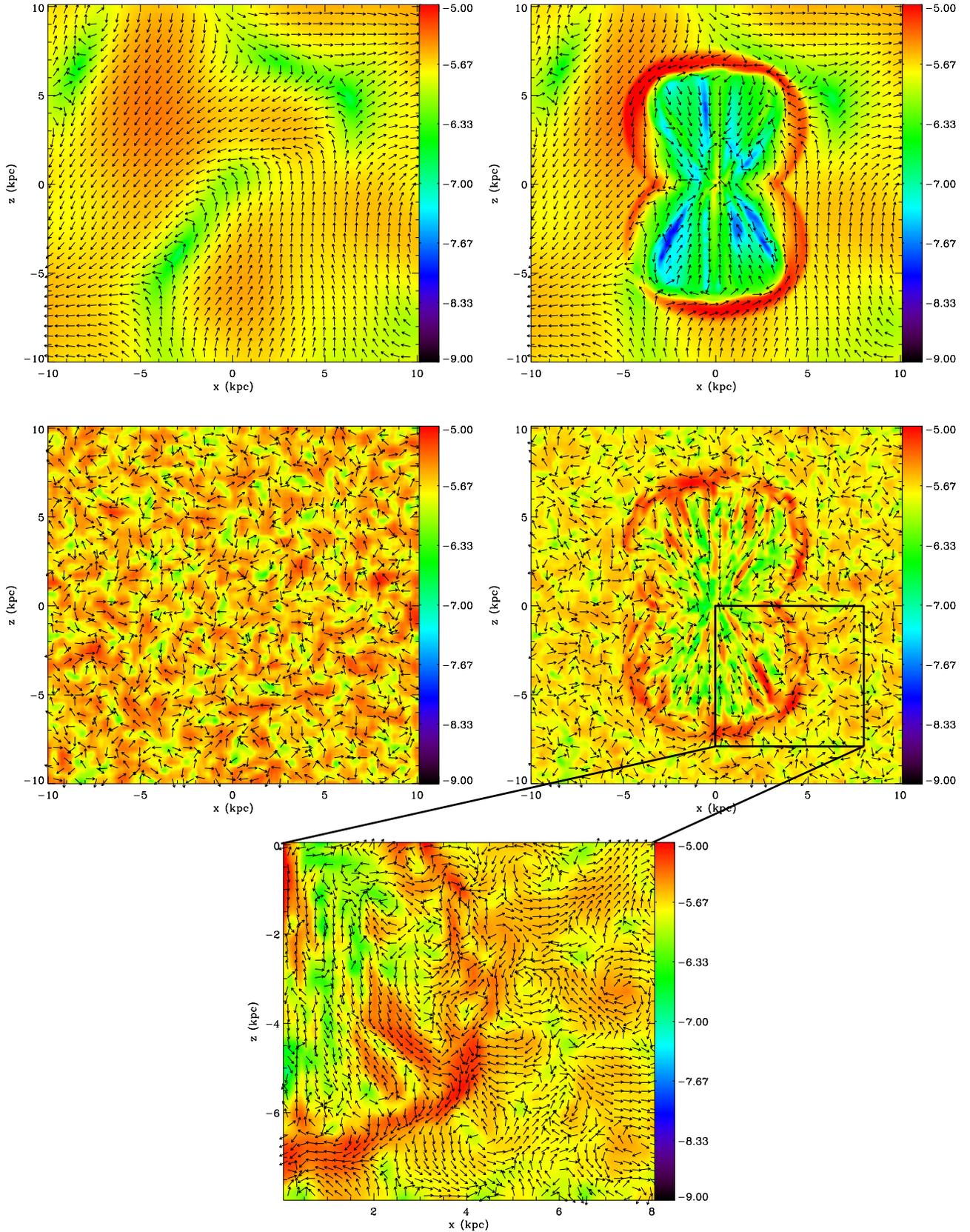} 
\caption{Slices in simulation coordinates of magnetic field magnitude (in logarithmic scale in units of $\mu$G) and directions (shown in arrows) at $t=0$ (left) and $t=t_{\rm Fermi}$ (right) for simulations without CR diffusion, Run $B$ (top row) and Run $G$ (second row). Run $B$ and $G$ has an initial magnetic coherence length of 9\ kpc and 1\ kpc, respectively. For clarity, a zoom-in image of Run $G$ at $t_{\rm Fermi}$ is shown in the bottom panel. Due to the effect of magnetic draping, the field is amplified and aligned with the bubble surface.}
\label{fig:magfield}
\end{center}
\end{figure*}

\subsection{Effects of CR Diffusion on Bubble Edges}
\label{sec:diffusion}

One of the most important characteristics of the observed {\it Fermi} bubbles is the sharp edges of their surface brightness. If CR diffusion is purely isotropic with a diffusion coefficient $\kappa_{\rm iso} \sim (3-5)\times 10^{28}\ {\rm cm}^2\ {\rm s}^{-1}$ \citep{Strong07}, the resulting CR distribution would have a much smoother edge than observed \citep{Guo11a}. The sharpness of bubble edges thus imply suppression of CR diffusion across the bubble surfaces \citep{Guo11a, Guo11b}. Using MHD simulations that self-consistently include effects of magnetic fields on CR diffusion, we will show that this suppression is likely due to anisotropic CR diffusion along magnetic field lines that drape around the {\it Fermi} bubbles. 

Figure \ref{fig:magfield} shows the magnetic field strength and topology at $t=0$ and $t=t_{\rm Fermi}$ for runs with different initial field coherence length, Run $B$ ($l_{\rm B}=9$\ kpc; top row) and Run $G$ ($l_{\rm B}=1$\ kpc; second row). A zoom-in image of Run $G$ at $t_{\rm Fermi}$ is shown in the bottom panel for clarity. For both simulations, the initial tangled field has an average strength of $\sim 1\ \mu$G, which corresponds to plasma $\beta=p_{\rm tot}/(B^2/8\pi) \sim 1$ in the ambient galactic halo (note that $p_{\rm cr}=0$ for the ambient gas), in agreement with observations \citep[e.g.,][]{Haverkorn12}. During the supersonic expansion of the {\it Fermi} bubbles, the ambient gas is compressed into shells with enhanced density and magnetic fields. Because of the magnetic draping effect \citep{Lyutikov06}, the field lines are stretched and compressed in the draping layer, causing field alignment with the bubble surface as well as amplification of field strength. The resulting alignment and amplification are {\it greater} when the direction of the field is initially {\it parallel} to the bubble surface (perpendicular to the direction of propagation). At $t=t_{\rm Fermi}$, the magnetic field within the shells is amplified to $\gtrsim 10\ \mu$G, comparable to the equipartition field strength inferred by radio polarization measurements \citep{Jones12}. Depending on the initial field strength and direction, the degree of field alignment and amplification at $t=t_{\rm Fermi}$ varies at different locations of the compressed shells, e.g., more on the northeast (upper left) side and less in the southeast (lower left) direction for Run $B$. Within the shells surrounding the CR bubbles, the orientation of magnetic fields generally lies preferentially along the bubble surfaces. The alignment and amplification is more pronounced for Run $B$ (larger initial coherence length) than Run $G$, which initially has a more randomly oriented field on small scales and the components perpendicular to the shock fronts are not draped effectively (e.g., the region at $x\sim 4$, $-4<z<-2$ in the zoom-in image in Figure \ref{fig:magfield}). As we will show later, the magnetic field topology at the bubble edges has a critical impact on the appearance of the simulated bubbles.  

Inside the bubbles, the magnetic field is more uniform for Run $B$ ($l_{\rm B}=9$\ kpc), with an average strength of $\sim 0.1\ \mu$G, whereas for Run $G$ ($l_{\rm B}=1$\ kpc), the magnetic field has a magnitude of $\sim 0.1-1\ \mu$G and is more filamentary. The enhanced field strength on the jet axis close to the GC comes from magnetic field injected by the AGN jets. The linear structure within the bubbles extending radially from the GC arises because of fields that are dragged along with the outwardly moving gas. We note that despite these enhancement, the magnetic field is dynamically unimportant for the expansion of the shocks and the CR bubbles, which is dominated by thermal pressure of the shocked gas and the CR pressure. In other words, the magnetic pressure is small compared to the total (gas plus CR) pressure, i.e., $\beta > 1$.   We find that except in the magnetized jets and some filaments where plasma $\beta \sim 1-5$, the magnetic field is largely dynamically negligible either in the CR bubbles ($\beta \gtrsim 50$) or within the compressed shells ($\beta \sim 10-50$). Consequently, the CR distribution at $t=t_{\rm Fermi}$ for these two runs looks almost identical to the purely hydrodynamic case Run $A$ (i.e., top panel of Figure \ref{fig:fiducial_slice}).         

\begin{figure*}[tp]
\begin{center}
\includegraphics[scale=0.7]{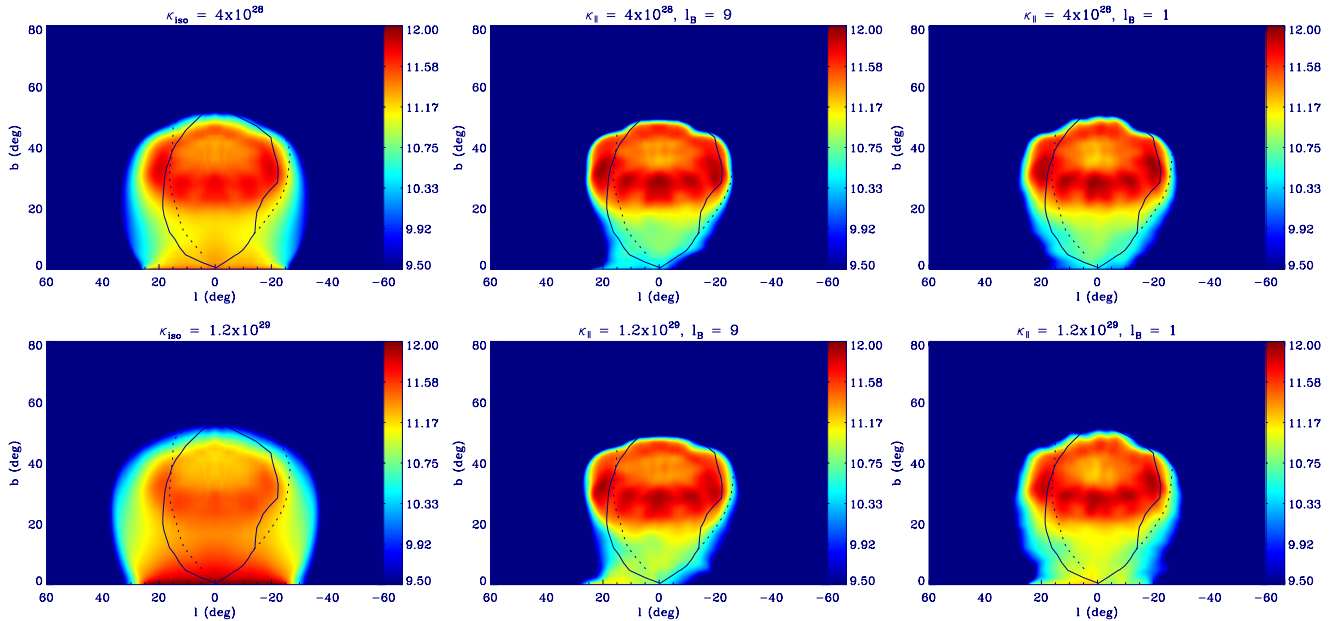} 
\caption{Maps of projected CR energy density in logarithmic scale in {\it cgs} units at $t=t_{\rm Fermi}$ for simulations including CR diffusion. Top and bottom rows show the runs with diffusion coefficient $4\times 10^{28}\ {\rm cm}^2\ {\rm s}^{-1}$ (Run $C$, $D$, and $I$) and $1.2\times 10^{29}\ {\rm cm}^2\ {\rm s}^{-1}$ (Run $E$, $F$, $K$), respectively. Columns from left to right are the cases of isotropic diffusion, anisotropic diffusion with $l_{\rm B}=9$\ kpc, and anisotropic diffusion with $l_{\rm B}=1$\ kpc, respectively. The solid and dashed lines are the observed northern and southern {\it Fermi} bubbles, respectively. Including isotropic diffusion results in a smoother CR distribution at the edges, whereas runs with  anisotropic diffusion produce sharper edges like the observed {\it Fermi} bubbles.}
\label{fig:ecr_diffusion}
\end{center}
\end{figure*}

The projected CR maps for simulations including CR diffusion are presented in Figure \ref{fig:ecr_diffusion}. Top and bottom panels are runs with diffusion coefficient $4\times 10^{28}\ {\rm cm}^2\ {\rm s}^{-1}$ and $1.2\times 10^{29}\ {\rm cm}^2\ {\rm s}^{-1}$, respectively. The panels from left to right show the cases of isotropic diffusion, anisotropic diffusion with $l_{\rm B}=9$\ kpc, and anisotropic diffusion with $l_{\rm B}=1$\ kpc, respectively. 
The isotropic cases with $l_{\rm B}=1$\ kpc (Run $H$ and Run $J$) are omitted here because the results are essentially identical to the corresponding $l_{\rm B}=9$\ kpc runs, because isotropic diffusion does not depend on the magnetic field geometry. 
Compared to the fiducial run without CR diffusion (Run $A$; Figure \ref{fig:ecrmap}), the cosmic rays in the isotropic diffusion cases (Run $C$ and Run $E$) have a smoother distribution because they are transported by CR advection and additional diffusion. Because of the dilution of CR energy density of the bubbles by diffusion, it takes a slightly longer time for the CR bubbles of $e_{\rm cr}\geq 10^{-11}\ {\rm erg\ cm}^{-3}
$ to reach the observed latitude, i.e., $t_{\rm Fermi}=1.22$\ Myr. Along the jet axis, the expansion of the CR bubbles is dominated by CR advection (recall that the gas velocities is $\sim 6000-7000\ {\rm km}\ {\rm s}^{-1}$ in the vertical direction), while in the lateral direction, the relative contribution of diffusion is greater with respect to advection (the gas velocities in the lateral direction is $\sim 2500\ {\rm km}\ {\rm s}^{-1}$). Consequently, the addition of CR diffusion has a more prominent effect laterally, yielding much fatter CR bubbles than the fiducial case. We also note that the diffusion transports cosmic rays to a larger distance than one would have estimated based on a simple dimensional relation, $l \sim \sqrt{\kappa t_{\it Fermi}}$, because of the large gradient of the CR energy density on the interface between the bubbles and the ambient gas. Therefore, the effect of CR diffusion is in fact quite substantial, and is easily visible in the figures.     

The most noticeable feature in the two cases with isotropic CR diffusion (Run $C$ and Run $E$; left panels in Figure \ref{fig:ecr_diffusion}) is the smooth transition of CR distribution at the edges, with CR energy density gradually declining outward. For $\kappa_{\rm iso}=1.2\times 10^{29}\ {\rm cm}^2\ {\rm s}^{-1}$, the smoothness due to diffusion can be easily seen, though may be exaggerated because such values of {\it isotropic} diffusion coefficients may be higher than observationally constrained \citep{Strong07}. However, even for an isotropic diffusion coefficient of $4\times 10^{28}\ {\rm cm}^2\ {\rm s}^{-1}$, as typically adopted in CR propagation models in the Galaxy \citep[e.g., GALPROP; ][]{Strong09}, the resulting CR distribution still has a smooth edge in contradiction to the sharp edges seen in the observed gamma-ray map. The observed sharpness of bubble edges thus  implies suppression of CR diffusion across the bubble surface.

The maps of projected CR energy density for runs with anisotropic diffusion are shown in the middle and right columns in Figure \ref{fig:ecr_diffusion}. In contrast to the cases of isotropic diffusion, the edges of the CR bubbles are much sharper, similar to the fiducial run where CR diffusion is not included (except perhaps Run $K$). The sharp edges are attributed to the combined effects of magnetic draping and anisotropic CR diffusion. Because of the small gyro-radius of relativistic particles, CR diffusion is anisotropic, i.e., cosmic rays may diffuse primarily along magnetic field lines with strongly suppressed cross-field diffusion. During the formation of the CR bubbles, since the magnetic field lines tend to align with the bubble surface due to magnetic draping (Figure \ref{fig:magfield}), CR diffusion occurs preferentially in the direction tangential to the bubble surface but is inhibited across the boundary. This suppression of CR diffusion across the bubble surface, which is a consequence of our MHD simulations that self-consistently include magnetic field and anisotropic CR diffusion, provides a natural explanation to the sharp edges of the observed {\it Fermi} bubbles.     

We note that the presence of the magnetic draping effect is manifested by comparing the CR distribution of Run $C$ ($\kappa_{\rm iso} = 4\times 10^{28}\ {\rm cm}^2\ {\rm s}^{-1}$) and Run $K$ ($\kappa_\parallel = 1.2\times 10^{29}\ {\rm cm}^2\ {\rm s}^{-1}$). Since Run $K$ has a tangled magnetic field on small scales, if the field is not aligned with the bubble surface but is completely random as it is initially, the result should correspond to an effective isotropic diffusion coefficient of $\sim 1/3$ of the parallel diffusion coefficient, i.e., $\kappa_{\rm iso} = 4\times 10^{28}\ {\rm cm}^2\ {\rm s}^{-1}$ as in Run $C$. The fact that the bubble edges are relatively sharper in Run $K$ than in Run $C$ indicates that, even though the magnetic field with a small coherence length may not be draped very effectively at some locations of the bubble surface, the magnetic draping effect {\it does} occur in a statistical sense (see the bottom panel of Figure \ref{fig:magfield}).     

Among the simulations with anisotropic diffusion (middle and right columns in Figure \ref{fig:ecr_diffusion}), there are a few differences depending on the diffusion coefficients and the structure of the underlying GMF. If the diffusion coefficient along the mean field is $\kappa_\parallel = 4\times 10^{28}\ {\rm cm}^2\ {\rm s}^{-1}$, the coherence length of the GMF has a minor effect on the appearance of the CR bubbles. On the other hand, if $\kappa_\parallel = 1.2\times 10^{29}\ {\rm cm}^2\ {\rm s}^{-1}$, the (possibly) largest diffusion coefficient in the {\it anisotropic} diffusion case permitted by current observational limits (see \S~\ref{sec:numerical}), the structure of the GMF may show observable features at the bubble edges. For example, in Run $K$ ($l_{\rm B}=1$\ kpc), where the magnetic field at the bubble surface is more randomly oriented (bottom right panel in Figure \ref{fig:magfield}), small-scale `wiggles' show up at the bubble edges due to fast diffusion along the field lines. The observed {\it Fermi} bubbles do not appear to have uneven surface at small scales. Therefore, by comparing our simulated CR image with the observation, we find that it is unlikely for our Galaxy to have fast diffusion ($\kappa_\parallel \sim 1.2\times 10^{29}\ {\rm cm}^2\ {\rm s}^{-1}$) in a GMF that has a small coherence length ($l_{\rm B} \sim 1$\ kpc). As for Run $F$ ($l_{\rm B}=9$\ kpc), the simulated northern CR bubble has a `tail' near the Galactic plane toward the east ($l\sim 10^\circ-20^\circ, b\sim 0^\circ-5^\circ$), which reflects the orientation of the underlying magnetic field at this location (Figure \ref{fig:magfield}, top right panel). Although our initial magnetic field is likely not the same as the actual GMF, our simulation results imply that the lack of absolute symmetry and even surface of the observed {\it Fermi} bubbles could be related to the structure of the underlying GMF. For instance, the slight extension or bend of the northern {\it Fermi} bubble at $l\sim -20^\circ, b\sim 30^\circ-40^\circ$ could possibly be the result of magnetic field that is oriented in the east-west (horizontal) direction and that was not effectively draped (see \S~\ref{sec:bend} for alternative explanations). Future observations of the GMF at the rims of the {\it Fermi} bubbles will help to verify whether this could be the case.

For the $l_{\rm B}=9$\ kpc cases shown in the middle column of Figure \ref{fig:ecr_diffusion}, because the cosmic rays diffuse mainly along the field lines that are draped around the bubble surface, even a diffusion coefficient as large as $\kappa_\parallel = 1.2\times 10^{29}\ {\rm cm}^2\ {\rm s}^{-1}$ is able to produce as sharp bubble edges as observed, just like the run with a smaller diffusion coefficient $\kappa_\parallel = 4\times 10^{28}\ {\rm cm}^2\ {\rm s}^{-1}$. This means that, if the GMF has a coherence length larger than the size of the {\it Fermi} bubbles, the CR diffusion coefficient parallel to the mean field cannot be well constrained solely by the sharpness of the bubble edges. This is similar to fossil radio bubbles observed in galaxy clusters, where the cosmic rays may be confined within the bubbles due to magnetic draping even when the parallel diffusivity coefficient is large \citep{Ruszkowski08}.    


\section{Discussion}
\label{sec:discussion}

\subsection{Parameter Dependencies and Constraints}
\label{sec:parameter}

In this section we comment on the parameter choices and dependencies. In the scenario where the {\it Fermi} bubbles are inflated by jets from the central AGN, \cite{Guo11a} has performed a detailed survey of jet parameters and derived plausible parameters by matching the shape of the observed bubbles. We found the same parameter dependencies of bubble morphology as theirs -- `thinner' (vertically-elongated) bubbles are produced by jets with a larger density contrast $\eta$, a smaller total energy density in the jets $e_{\rm j}+e_{\rm jcr}$ (which dominates the lateral expansion of the bubbles; also note that $B^2/8\pi$ is negligible), a larger jet velocity $v_{\rm jet}$, a smaller jet width $R_{\rm jet}$, or a longer jet duration $t_{\rm jet}$. As discussed in \cite{Guo11a}, these parameters are essentially degenerate in reproducing the observed bubble shape. Moreover, a number of jet parameters derived from their simulations scale with the assumed normalization of the initial gas density profile ($n_{\rm e0}$), including $\rho_{\rm j}$, $e_{\rm j}$, $e_{\rm jcr}$, and thus the total jet power $P_{\rm jet}$ and injected energy $E_{\rm jet}$. 

In order to reduce some of the parameter degeneracies and improve the estimations of jet parameters, in this study we introduce several additional observational constraints. Firstly, for the hot gaseous halo we choose the initial temperature $T_0$ and the normalization constant $n_{\rm e0}$ to match the observed gas density profile of \cite{Miller12} (\S~\ref{sec:galmodel}). This is particularly important because the supersonically expanding bubbles are partially confined by the pressure of the ambient gas, and thus the final shape of the bubbles depends on the chosen initial gas profiles. However, we note that there are still large uncertainties in the current constraints on the observed profiles of the hot gaseous halo \citep{Fang06, Miller12}. Moreover, these measurements only provide information about the gas profile as observed {\it today} (especially after it is already modified by the AGN jet event in the inner few kpc), rather than the initial condition about $\sim 1$\ Myr ago when the AGN jets were first launched. Therefore, though our simulated gas density profile at $t_{\rm Fermi}$ is within the observed limits, there are still uncertainties in the assumed initial hot gas distribution in the vicinity of the SMBH, and hence also in the estimated jet power and total energy. 

Secondly, observations of X-ray absorption line ratios of OVIII to OVII put a constraint on the temperature of the shocked gas inside the {\it Fermi} bubbles to be $T\gtrsim 10^7$\ K \citep{Miller12}. We find that slower and wider jets are required to fulfill the constraints from the observed line ratios (\S~\ref{sec:jet}), implying that the relativistic jets from the AGN may have slowed down a little by interacting with the interstellar medium during propagation from pc to kpc scales \citep[e.g.,][]{Middelberg04}. Since these modifications would make the bubbles much `fatter', we had to either increase $\rho_{\rm j}$, $t_{\rm jet}$, or decrease $e_{\rm j}+e_{\rm jcr}$ in order to reproduce the observed bubble shape. However, $\rho_{\rm j}$ and $t_{\rm jet}$ cannot be arbitrarily large, otherwise the electron number density at $t_{\rm Fermi}$ inside the bubbles would be too high and would violate the limb-brightened X-ray observations by {\it ROSAT} \citep{Snowden97, Su10}. Therefore, the total energy density in the jets, $e_{\rm j}+e_{\rm jcr}$, becomes the primary variable used to fit the observed shape of the bubbles. 

We find that these observations set very stringent constraints on the allowed parameter space for the jets; for a given initial gas profile, varying any of the jet parameters could easily violate one of the observational requirements. It is remarkable that after applying these constraints, the permitted parameters are able to {\it simultaneously} reproduce the many characteristics of the observed {\it Fermi} bubbles, including the short age (thus consistent with the constraint from the IC cooling timescale), the absence of large-scale instabilities, and the limb-brightened CR distribution needed for uniform gamma-ray surface brightness. The only remaining degeneracy is between the injected thermal energy density $e_{\rm j}$ and CR energy density $e_{\rm jcr}$, as their sum affects the pressure contrast with respect to the ambient medium and helps drive the expansion of the bubbles (though dominated by the ram pressure of the jets). This degeneracy can in principle be broken by detailed comparisons with the radio and gamma-ray observations.      

\subsection{Bends of the Bubbles}
\label{sec:bend}

The properties of our simulated CR bubbles are in broad agreement with the observed ones. However, the observed {\it Fermi} bubbles are not completely symmetric about the GC. Both the northern and southern bubbles are slightly bent to the {\it west} (negative longitude). In this section we discuss possible causes of this asymmetry.     

Up to now we have only considered the simplest case of bipolar jets that are normal to the Galactic plane. However, AGN jets, particularly on kpc scales, are not necessarily aligned with the rotational axis of the galactic disks, and such misalignments are observed frequently in other galaxies \citep[e.g.,][]{Middelberg04, Giroletti05, Gallimore06, Kharb10}, and also recently found in the Milky Way \citep{Su12}. This misalignment could either arise from an intrinsic effect, such as jet precessions \citep[e.g.,][]{Falceta-Goncalves10}, or could be due to jet interactions with the surrounding medium as the jet travels from pc to kpc scales (e.g., Fanaroff-Riley type I radio sources). Forming the {\it Fermi} bubbles by intrinsically tilted jets is appealing and is well-motivated by the slanted gamma-ray jets recently discovered near the GC \citep{Su12}. However, in this case, the two jets would point at opposite directions at a $180^\circ$ angle. They would potentially form two bubbles, with the northern one bending to the west, and the southern one to the east, inconsistent with the {\it Fermi} bubbles that both bend to the west. This argument, together with the fact that the gamma-ray jets have a spectrum at $<1$\ GeV different from the {\it Fermi} bubbles \citep{Su12}, suggest that the bubbles probably do not originate from the newly-discovered gamma-ray jets. 

Bending of radio jets in some cases can be explained by external ram pressure as galaxies move through an intracluster gas \citep[e.g.,][]{Begelman79, Balsara92}. It is known that the Milky Way moves in the intragroup medium (IGM). If the ram pressure of the wind generated by such motion is sufficient, it can potentially `blow' the bubbles toward the same direction. Assuming the {\it Fermi} bubbles are tilted by $\alpha$ degrees from the rotational axis of the Galaxy, the required ram pressure from the wind \citep{Burns79} is $P_{\rm ram, wind} = P_{\rm ram, jet} \tan{\alpha}$, where $P_{\rm ram, jet}=\rho_{\rm j} v_{\rm jet}^2$ is the ram pressure of the jets. Taking $\alpha=10^\circ$ and our fiducial jet parameters, $P_{\rm ram,jet}=1.1\times 10^{-7}\ {\rm dyne\ cm}^{-2}$, we obtain $P_{\rm ram,wind}\sim 2\times 10^{-8}\ {\rm dyne\ cm}^{-2}$, corresponding to a wind of density $\rho_{\rm wind}\sim 4\times 10^{-24}\ {\rm g\ cm}^{-3}$ and velocity $v_{\rm wind}\sim 750\ {\rm km\ s}^{-1}$. These values are unreasonably large for estimated properties of the Local Group \citep[e.g.,][]{McConnachie07, Cox08}. Moreover, such strong wind would have dominated over the gravitational restoring force \citep{McCarthy08} and stripped a significant amount of Milky Way gas away. Therefore, given reasonable values of the IGM, the resulted wind is unlikely to have sufficient impacts on the bubbles to the desired degree.     

The required ram pressure, instead of being external, could possibly come from the region in the SMBH vicinity. If the SMBH moves relative to the hot gaseous halo, it can effectively experience a wind and the ram pressure over time might affect the direction of the jets. However, little motion of the SMBH at the GC is detected based on proper motion measurements of the Sgr A* \citep{Reid05}. 

Another possibility is the ram pressure from supernova (SN) explosions. It is well known that the Sgr A* is surrounded by nuclear star clusters, among which exists a population of massive young stars that formed $\sim 6$\ Myr ago in the inner 0.5 pc from the SMBH \citep{Paumard06}. These massive stars typically have main-sequence lifetimes of a few Myr, and hence it is possible that one of them exploded as a SN during or sometime before the active phase of the AGN jets. In such a case, assuming a compressed interstellar medium of density $\rho \sim 10^{-25}\ {\rm g\ cm}^{-3}$ and a typical SN shock velocity of $v \sim 10^4\ {\rm km\ s}^{-1}$, the ram pressure generated by the SN would be sufficient to bend the jets. One could more accurately calculate the ram pressure as a function of time, the ambient gas density and the initial energy of explosion from, for example, the Sedov solution \citep{Sedov46}. We found that for a typical SN energy of $10^{51}\ {\rm erg}$ and ambient density of $10^{-25}\ {\rm g\ cm}^{-3}$, the duration within which the SN can effectively bend the jets by more than $10^\circ$ is $\sim 10^4$\ yr, which is about one tenth of the duration of our jets. 

\begin{figure}[tp]
\begin{center}
\includegraphics[scale=1.0]{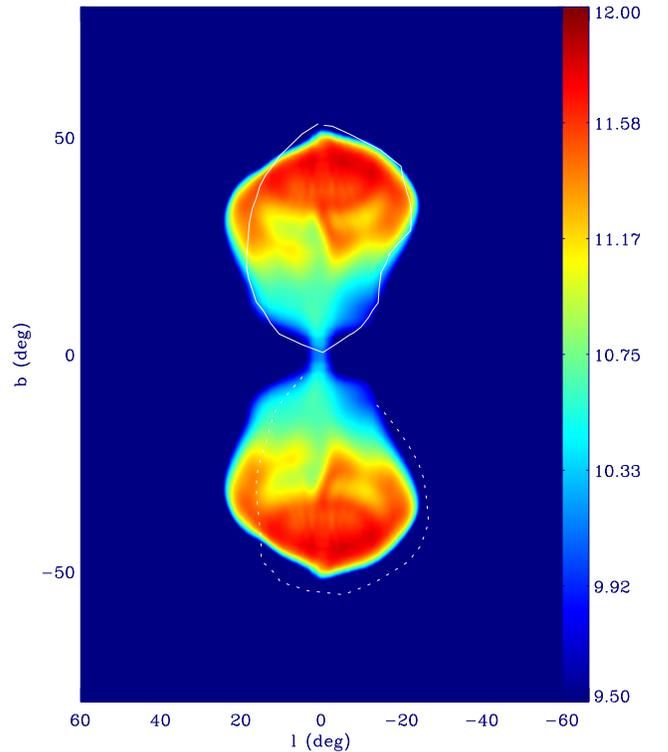} 
\caption{Projected map of CR energy density in logarithmic scale in {\it cgs} units for the case where both CR jets are tilted to the {\it east} by $10^\circ$ (possibly due to SN ram pressure; see texts) from the rotational axis of the Galaxy for $0 < t < 0.1\ t_{\rm jet}$, then returned to the normal direction for $0.1\ t_{\rm jet} < t < t_{\rm jet}$. Solid and dotted lines show the edges of the observed northern and southern {\it Fermi} bubbles, respectively. } 
\label{fig:tilt}
\end{center}
\end{figure}

In order to see whether such SN event could cause the observed bends of the {\it Fermi} bubbles of approximately the right morphology, we did a numerical experiment in which the northern and southern CR jets are both tilted by $10^\circ$ to the negative $x$-axis (positive Galactic longitude) from the rotational axis of the Galaxy for $0 < t < 0.1\ t_{\rm jet}$, and then returned to the normal axis for $0.1\ t_{\rm jet} < t < t_{\rm jet}$. The projected CR energy density at $t_{\rm Fermi}=1.4$\ Myr is shown in Figure \ref{fig:tilt}. As can be seen, the `left'-bent jets result in a fatter CR distribution in the east, causing asymmetries about the rotational axis. The simulated bubbles have a remarkable resemblance to the observed bubble morphology, considering the simple assumptions and models employed. This suggests that bent AGN jets acting even for a short period of time, possibly due to ram pressure from a SN explosion that occurred near the SMBH, could plausibly cause the slight bends of the observed {\it Fermi} bubbles. 

As mentioned in \S~\ref{sec:diffusion}, the asymmetry of the observed bubbles could also be a result of large CR diffusivity combined with magnetic fields orienting perpendicular to the bubble surface. However, a special field configuration would be required for both bubbles to bend toward the same direction. Improved observations of the field geometry near the bubble edges will help to verify this possibility.     

\subsection{Comparison with ROSAT X-ray Map}
\label{sec:xray}

The {\it ROSAT} X-ray 1.5\ keV map \citep{Snowden97} has revealed enhanced emission surrounding the northern {\it Fermi} bubble, which is likely produced by bremsstrahlung emission from shocked gas during bubble formation \citep{Su10}. The observed X-ray emission is limb-brightened, suggesting that the bubbles are hot and {\it underdense}. As discussed in \S~\ref{sec:parameter}, this provides an important constraint on the thermal content of the AGN jets, i.e., the jets cannot be too `heavy', otherwise the bubbles would be too bright on the X-ray map as they would be filled with large amounts of thermal gas. In \S~\ref{sec:morphology} we have shown that our simulated bubbles are indeed underdense and hot (see bottom panel of Figure \ref{fig:fiducial_slice}). However, the projections from the 3D distribution onto the 2D maps may be nontrivial, affecting the X-ray intensity distribution and location of the shocks, for instance. In order to show general consistency with the observed X-ray images, we make a simulated X-ray map by projecting the bremsstrahlung emissivity computed from the density and temperature of the simulated gas. 

For the simulated X-ray map, the X-ray emissivity in an energy range $1.4-1.6$\ keV is calculated using the MEKAL model \citep{Mewe85, Kaastra93, Liedahl95} implemented in the utility XSPEC \citep{Arnaud96}, assuming solar metallicity. Note that the observed X-ray emission is contributed by {\it all} the gas in the Milky Way halo, which likely extends to a radius of $\sim 250$\ kpc \citep{Blitz00, Grcevich09}, much bigger than our simulation box. Therefore, we first compute the X-ray emissivity from the simulated gas within a radius of $25$\ kpc away from the GC. Then, beyond $25$\ kpc the gas is assumed to be isothermal with $T=2\times 10^6$\ K and follows out to a radius of 250\ kpc the observed density profile of \cite{Miller12} with an electron number density floor, $n_{\rm e,floor}$. Studies of ram pressure stripping of dwarf spheroidal galaxies orbiting the Milky Way give constraints to the value of $n_{\rm e,floor}$ to be around $2.4\times 10^{-5} - 2.5\times 10^{-4}\ {\rm cm}^{-3}$ \citep{Blitz00, Grcevich09}. Here we choose $n_{\rm e,floor} = 8.0\times 10^{-5}\ {\rm cm}^{-3}$ so that the OVII column density computed from this extended gaseous halo is consistent with the observed values \citep{Miller12}. We show the results from the same run as in the previous section (with a short period of jet bending) as it produces bubbles with morphology closest to that observed.  

\begin{figure}[tp]
\begin{center}
\includegraphics[scale=1.0]{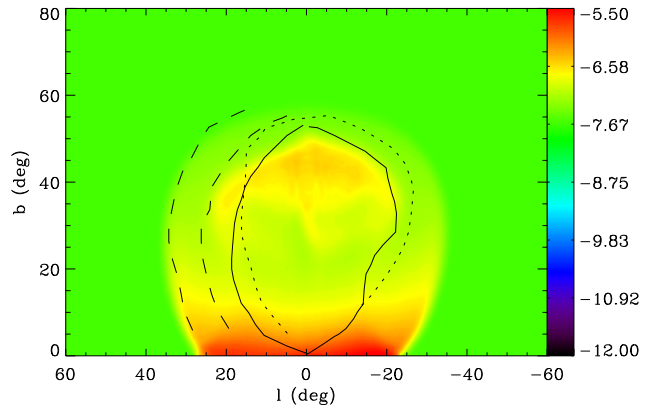} 
\caption{Simulated X-ray map at 1.5\ keV (see texts for definition) for the same run as Figure \ref{fig:tilt}. The solid and dotted lines show the surfaces of the observed northern and southern {\it Fermi} bubbles, respectively. The dash lines are the inner and outer northern arcs observed by the {\it ROSAT} X-ray satellite.} 
\label{fig:xraymap}
\end{center}
\end{figure}

The projected X-ray emissivity at 1.5\ keV is shown in Figure \ref{fig:xraymap}, overplotted with the observed contours of the {\it Fermi} bubbles and arc features identified by {\it ROSAT}. A region more extended than the bubbles is X-ray bright compared to the ambient medium because the gas is compressed and heated by the shocks during the bubble expansion. The projected location of the shock fronts is in excellent agreement with the observed outer northern arc feature embracing the bubbles. The projected map has a much smoother distribution than one would have naively expected from a 2D slice that shows a clear cavity within the shock compressed shells (as in the bottom panel of Figure \ref{fig:fiducial_slice}). This is because the lines of sight toward the bubble interior also pass through the closer and further ends of the shells, and thus the change in projected X-ray emissivity with sight lines moving toward the shell edges is rather continuous. The projected X-ray emission inside the bubbles is dimmer and gradually increases toward the edges of the shocked gas, consistent with the limb-brightened distribution on the observed X-ray map. We note that the approach we use to produce the X-ray map is rather simplistic; simulated X-ray observations including stochastic photon generation and instrumental responses for different bands are required to make more detailed comparisons with the observational data. However, the general success in reproducing the observed X-ray features provides a strong support for the hypothesis that both the {\it Fermi} bubbles and the {\it ROSAT} X-ray features originate from the same episode of jet activity from the central SMBH. 

\section{Conclusions}
\label{sec:conclusion}

The {\it Fermi Gamma-ray Space Telescope} has recently revealed two large bubbles extending $\sim 50^\circ$ above and below the GC, with a width of $\sim 40^\circ$ in longitude. The northern and southern bubbles are nearly symmetric about the Galactic plane, with only slight bends to the west. Their spectrum is hard and has no significant variations within the bubbles. The gamma-ray surface brightness is quite uniform with sharp edges at the boundaries.  Besides, the gamma-ray bubbles are spatially correlated with features in the {\it ROSAT} X-ray map and the hard-spectrum WMAP haze. It is challenging to explain all these observed properties, for example, by SN shocks in the Galactic disk, or by dark matter annihilation \citep{Su10}.

The symmetry about the GC of the two bubbles, the bilobular shape, and their similar hard spectrum strongly suggest the bubbles originated from an energetic event in the GC within the past few Myr. Forming the bubbles and their observed shape by a past jet activity of the central SMBH is shown to be plausible by \cite{Guo11a} using 2D hydrodynamic simulations. 
They also found that the sharpness of bubble edges requires suppression of CR diffusion across the bubble surface, which may be related to the interplay between CR diffusion and magnetic fields. However, their simulated bubbles reveal several discrepancies with the observations, including the rippled surface due to hydrodynamic instabilities, and the inferred centrally-brightened surface brightness. To this end, shear viscosity is invoked in their second paper \citep{Guo11b} to alleviate these problems.  

In this study, we investigate the jet scenario using a set of 3D MHD simulations that self-consistently include the effects of magnetic fields and anisotropic (field-aligned) CR diffusion, as well as dynamical interaction between the thermal gas and cosmic rays. The simulations are performed using a new CR module in the FLASH code, which we have implemented and tested (see Appendix \ref{appendix}). We summarize our findings as follows.

1.\ The effect of projection of the 3D bubbles has a significant impact on the estimation of the bubble formation time. Because the widths of the bubbles occupy a non-negligible fraction of the distance from the Sun to the GC, for the 3D CR bubbles to project onto a Galactic latitude of $50^\circ$, their vertical dimension only needs to be $\sim 6$\ kpc, instead of $\sim 10$\ kpc as previously thought. The projection effect results in a much shorter bubble formation time ($\sim 1.2$\ Myr for our fiducial model) than in previous estimations.  

2.\ If the observed gamma-ray emission is produced by IC scattering of CR electrons by the ISRF, the IC cooling time of high-energy ($\sim 100$\ GeV) electrons gives a stringent upper limit to the bubble ages to be within a few Myr. This constraint is naturally satisfied by the 'young' bubbles revealed by our 3D simulations.  

3.\ Because of the short ages of our simulated bubbles, there is no sufficient time for large-scale hydrodynamic instabilities to grow. This alone explains why the observed {\it Fermi} bubbles have a rather smooth surface (as opposed to rippled), and there is no need to invoke other mechanisms (e.g., viscosity or magnetic fields) to explain the suppression of the instabilities.  

4.\ Our jet parameters, which are determined by various observational constraints (see \S~\ref{sec:parameter}), produce an edge-enhanced CR distribution that, when projected onto the plane of the sky, result in a roughly flat surface brightness distribution as a function of the Galactic longitude. The projected CR intensity increases with the Galactic latitude.
The apparent discrepancy with the flat gamma-ray surface brightness may be explained by the decaying ISRF away from the Galactic plane, if the cosmic rays are primarily leptonic, or by the large uncertainties in the observed data at lower latitudes due to background subtraction.

5.\ The sharp edges of the {\it Fermi} bubbles are reproduced by self-consistently including anisotropic CR diffusion along magnetic field lines that drape around the bubble surface during the bubble expansion.  

6.\ The causes of the slight bends of the {\it Fermi} bubbles are discussed. Possible explanations include jet bending due to ram pressure from SN near the SMBH, and fast CR diffusion along magnetic field lines that initially lie perpendicular to the bubble surface (see \S~\ref{sec:bend}).

7.\ The projected X-ray bremsstrahlung emissivity of the shocked gas due to bubble expansion successfully reproduces the location and limb-brightened property of the X-ray features surrounding the {\it Fermi} bubbles observed with {\it ROSAT}. This provides evidence that the {\it ROSAT} X-ray features originated from the same AGN jet activity episode as the {\it Fermi} bubbles.    

The properties of our simulated bubbles are in broad agreement with the key features of the {\it Fermi} bubbles, strengthening the case that the bubbles are created by a recent AGN jet activity at the GC. Such event is analogous to bipolar radio bubbles and X-ray cavities associated with the nuclei of massive galaxies \citep{McNamara07}. {\it Fermi} cavities are most likely created by jets from the central SMBH, e.g., as those observed in distant galaxies (e.g., M87). Though the SMBH at the center of the Milky Way is currently in a quiescent state \citep[see, however, a pair of gamma-ray jets recently discovered by][]{Su12}, the detection of the {\it Fermi} bubbles, together with indications of past activity \citep{Koyama96, Revnivtsev04}, suggest that the central SMBH may have regularly undergone cycles of jet activity, 
or even was at a much more active state for the past 1-10\ Myr than now \citep{Totani06}.
The energy input from these jets may have altered the properties of the multiphase gas in the Galactic bulge and halo, suppressed the accretion process, and regulated the co-evolution of the SMBH and the stellar bulge in the past. Moreover, these AGN jets may be a significant source of the cosmic rays in the Galactic halo \citep{Guo11a}. 
We note that the total energy required to inflate the bubbles depends on how cuspy the central density profile is; for more cored profiles, the energy of the jet will be lower.
The proximity to the SMBH in the GC offers a unique opportunity to study these processes in detail, especially in gamma rays due to the limited sensitivity and resolution of gamma-ray observations.      

The {\it Fermi} bubbles have been identified with a unique source of cosmic rays near the Galactic center. If these cosmic rays are leptonic, i.e., composed of electrons and positrons, they could possibly annihilate and produce a characteristic line emission at 511\ keV. One may ask whether this is linked to the enhanced 511\ keV radiation observed in the Galactic bulge \citep[see a review by][]{Prantzos11}. However, we find that this is unlikely to be the case because the cosmic rays inferred from the observed bubble emission are above GeV, and for these high-energy electrons and positrons the expected line flux at 511\ keV is extremely small due to small annihilation cross sections. In order to have significant line emission, the leptons would need to be cooled down to MeV ranges. However, little CR cooling is expected from the hard spectrum of the observed bubbles. Moreover, if there were significant cooling, the in-flight annihilation of high-energy cosmic rays would over-produce the emission in the MeV range \citep[see Figure 6 of ][]{Prantzos11}.

Finally, we note that in our simulations we have only considered the ensemble of cosmic rays and neglected distinctions between different CR species and energies. However, many physical processes depend on the types of CR particles under consideration and their energy spectra, such as CR production and reacceleration in shocks, and energy losses due to adiabatic expansion, synchrotron cooling and IC emission. The amount of cosmic rays that could be accelerated inside the shocks could be roughly estimated based on shock properties in our simulations and an assumed model for diffusive shock acceleration \citep{Ensslin07}. We find that the shock-accelerated cosmic rays mostly have energies around a few MeV; the number density of cosmic rays above 1\ GeV, which can produce gamma-ray emission in the observed energy range of the bubbles, is only a few percent of the simulated CR number density (assuming a spectral index of -2 and energy range of 0.1-1000\ GeV). Although the low-energy cosmic rays could possibly somewhat contribute to the total pressure and help drive the dynamical evolution of the bubbles, an effect we will investigate in detail in the future, their contribution to the bubble emission is expected to be small. Interestingly, this is supported by the fact that the observed gamma-ray emission of the {\it Fermi} bubbles does not extend to the locations of the X-ray arc features.

Energy losses due to adiabatic expansion, synchrotron and IC cooling could also cause evolution of CR energy spectrum during the bubble expansion. During adiabatic expansion, although the process is energy independent and does not alter the shape of the CR spectrum, the cosmic rays can lose energy and shift the spectrum to lower energies. The shift in CR energies is directly proportional to the change in the CR energy density. The ratio of the CR energy density in the edge-brightened region of the bubble at the current epoch to the energy density in the early stage in the bubble evolution is around $\sim 100$. Therefore, taking into account only adiabatic contribution, the CR energies in the initial state were $\sim 100$ larger than the corresponding energies at $t_{\rm Fermi}$. This implies that CR cooling times due to synchrotron and IC scattering are shorter in the early stages in the bubble evolution. The actual spectral shape at $t_{\rm Fermi}$ will however also depend on other factors that may hide or offset these losses. These factors include: jet speed, magnetic field in the bubble, and possible CR reacceleration close to the GC. These processes will have to be accounted for with detailed modeling of the CR spectra. However, the fact that the simulated bubble ages are quite short helps to reconcile the simulations with the spectral constraints from observations.

Furthermore, as discussed in \S~\ref{sec:parameter}, our current model is degenerate with respect to the internal energy density and CR energy density in the jets. A comparison of the simulated spectra in the gamma-ray and/or radio bands with the observational data should allow to break this degeneracy. This technique may also potentially offer a unique way to constrain the contents of the AGN bubbles and is likely to be superior to the constraints from AGN bubbles in galaxy clusters because of the proximity of the GC and the availability of additional spatially resolved data from {\it Fermi}.   








\acknowledgments

The authors would like to thank Greg Dobler, S.\ Peng Oh, and Yen-Hsiang Lin for helpful discussions. We thank Fabian Heitsch for help in the early stages of the project, and Min-Su Shin for discussions during the final stages of this work. We thank Joel Bregman and Matt Miller for valuable comments and for sharing us the results of their work prior to publication. HYKY and MR acknowledge the NSF grant NSF 1008454. PMR acknowledges support from NASA under the Fermi Guest Investigator Program (NNX10AO78G). EZ acknowledges support from the NSF grant (NSF AST0903900). DL acknowledges support from the DOE (contract number: 0J-30381-0005A) and the NSF (PHY-0903997). FLASH was developed in part by the DOE NNSA ASC- and DOE Office of Science ASCR-supported Flash Center for Computational Science at the University of Chicago.


\bibliography{fermi}


\appendix

\section{A.\ Numerical Implementations and Tests for Cosmic Rays}
\label{appendix}

In this study, we introduce a new module in the FLASH code for simulating cosmic rays as a second fluid. A brief summary of the code and the directionally unsplit staggered mesh (USM) solver, as well as the general assumptions and equations for simulating CR advection and diffusion, are already discussed in \S~\ref{sec:numerical}. In this appendix, we first describe the implementation details of how we incorporate cosmic rays into the USM solver, and then provide results of several numerical tests of the new CR module.   

The MHD equations including CR advection, dynamical coupling of thermal gas and cosmic rays, and anisotropic CR diffusion, are listed in Eq.\ 1-6 in \S~\ref{sec:numerical}. The mass, momentum, total energy, and induction equations are solved by the Piecewise-Parabolic Method (PPM; \cite{Colella84}) in the USM solver \citep{Lee09, Lee12}. In the absence of CR diffusion, the only modification is that the pressure contains an additional contribution from the cosmic rays, i.e., $p_{\rm tot}=p_{\rm th}+p_{\rm cr}+p_{\rm B}=(\gamma-1)e_{\rm th}+(\gamma_{\rm cr}-1)e_{\rm cr}+B^2/8\pi$, where $e_{\rm th}$ and $e_{\rm cr}$ and the internal energy densities, and $\gamma$ and $\gamma_{\rm cr}$ are the adiabatic indices of the thermal gas and cosmic rays, respectively. Accordingly, the sound speed of the combined fluid of gas and cosmic rays, which is used when solving the Riemann problem in the PPM method as well as when computing the hydrodynamic timestep, is replaced by the {\it effective} sound speed \citep{Miniati07}, 
\begin{equation}
c_s^\prime = \sqrt{\frac{\gamma p_{\rm th} + \gamma_{\rm cr} p_{\rm cr}}{\rho}}.
\end{equation}
We note that when the left and right states are reconstructed at cell boundaries in the PPM method, it is crucial that the CR energy is also reconstructed in the same way as other gas variables (i.e., to the same order and through the same slope limiters) in order to avoid spurious oscillations around discontinuities (e.g., Figure \ref{fig:shocktube}).  

The energy of cosmic rays is evolved from Eq.\ \ref{eq:ecr}, in which the advection term is advanced using the mass scalars in FLASH that passively evolve with the thermal gas, and then the source term $-p_{\rm cr}\nabla\cdot {\bm v}$ is updated in the discretized form (in 2D) as
\begin{equation}
-p_{{\rm cr}, i,j}^* \cdot \left[ \frac{v_{i+1/2,j}^{n+1/2} - v_{i-1/2,j}^{n+1/2}}{\Delta x} + 
\frac{v_{i,j+1/2}^{n+1/2} - v_{i,j-1/2}^{n+1/2}}{\Delta y} \right],
\end{equation}
where $\Delta x$ and $\Delta y$ are the sizes of grid cells, $v_{i \pm 1/2,j}^{n+1/2}$ and $v_{i,j \pm 1/2}^{n+1/2}$ are velocities at cell boundaries reconstructed by the PPM method, and $p_{{\rm cr}, i,j}^*$ is the averaged CR pressure before and after advection. Extension to 3D is straightforward, and we skip it here for the sake of brevity. 

The anisotropic CR diffusion terms in the total energy and CR energy equations are updated explicitly using fluxes computed from the conservative `centered asymmetric differencing' scheme \citep{Parrish05, Sharma07}. The updated energy (for either the total or CR energy) in 2D is 
\begin{equation}
e_{i,j}^{n+1} = e_{i,j}^n - \Delta t \left[ 
\frac{F^n_{i+1/2,j}-F^n_{i-1/2,j}}{\Delta x} +
\frac{F^n_{i,j+1/2}-F^n_{i,j-1/2}}{\Delta y} \right],
\end{equation}
The simulation timestep $\Delta t$ is required to satisfy both the CFL criterion and the stability condition for CR diffusion, 
\begin{equation}
\Delta t \leq \frac{{\rm min}[\Delta x^2,\Delta y^2]}{2(\kappa_\parallel+\kappa_\perp)}.
\end{equation}
Using the asymmetric method, the $x$-flux (similar for the $y$-fluxes) at the cell face $i+1/2$ at time $n$ due to anisotropic CR diffusion (Eq.\ \ref{eq:anisotropic}) is given by 
\begin{eqnarray}
F^n_{i+1/2,j} = -(\overline{\kappa_\parallel} - \overline{\kappa_\perp}) b_x 
&& \left[ b_x \frac{\partial e_{\rm cr}}{\partial x} + 
\overline{b_y} \overline{\frac{\partial e_{\rm cr}}{\partial y}}\right] 
- \overline{\kappa_\perp} \frac{\partial e_{\rm cr}}{\partial x},\\
b_x = && b^n_{x,i+1/2,j}, \\
\frac{\partial e_{\rm cr}}{\partial x} = && \frac{e_{{\rm cr},i+1,j} - e_{{\rm cr},i,j}}{\Delta x}, \\
\overline{b_y} = (b_{y,i,j-1/2} + b_{y,i+1,j-1/2} && + b_{y,i,j+1/2} + b_{y,i+1,j+1/2})/4,\\
\frac{2}{\overline{\kappa_\parallel}} = \frac{1}{\kappa_{\parallel,i,j}} + \frac{1}{\kappa_{\parallel,i+1,j}}, && \ 
\frac{2}{\overline{\kappa}_\perp} = \frac{1}{\kappa_{\perp,i,j}} + \frac{1}{\kappa_{\perp,i+1,j}}, 
\end{eqnarray}
and the transverse flux is sloped-limited in order to avoid negative CR energy densities when there is a large gradient \citep{Sharma07},
\begin{equation}
\overline{\frac{\partial e_{\rm cr}}{\partial y}} = L \left\{ 
L \left[ \left. \frac{\partial e_{\rm cr}}{\partial y} \right|_{i,j-1/2}, 
\left. \frac{\partial e_{\rm cr}}{\partial y} \right|_{i,j+1/2} \right],
L \left[ \left. \frac{\partial e_{\rm cr}}{\partial y} \right|_{i+1,j-1/2}, 
\left. \frac{\partial e_{\rm cr}}{\partial y} \right|_{i+1,j+1/2} \right] \right\},
\end{equation}
where $L$ is a slope limiter such as minmod, van Leer, or Monotonized Central (MC) limiter. We adopted the MC limiter, defined as
\begin{eqnarray}
{\rm MC}(a,b)= {\rm minmod} \left[ 2 {\rm minmod}(a,b),\frac{a+b}{2} \right],\\
{\rm minmod}(a,b) = \left\{ 
\begin{array}{ll}
{\rm min}(a,b) & \text{if $a,b>0$}, \\
{\rm max}(a,b) & \text{if $a,b<0$}, \\
0               & \text{if $ab \leq 0$}.\\
\end{array} \right.
\end{eqnarray}

\subsection{A.1\ Test for CR Advection}
\label{sec:adv}

First we perform a simple 2D test where cosmic rays are advected along the diagonal direction. 
The simulation domain of size $1\times 1$ is initially filled with a uniform thermal gas of density $\rho_0$ and pressure $p_0$, moving with the velocity $(v_{x0},v_{y0})$. On top of the thermal gas we placed a Gaussian overdensity of cosmic rays, 
\begin{equation}
e_{\rm cr} = e_{\rm cr,0} (1+ \exp[-0.5(r/r_0)^2] ). \label{eq:gauecr}
\end{equation}
The following parameters are adopted: $\rho_0=1$, $p_0=1$, $e_{\rm cr,0}=0.1$, $r_0=0.05$, $v_{x0}=1$, and $v_{y0}=1$. Since the purpose of this test is only to verify the implementation of CR advection, the CR pressure is set to be much smaller than the thermal gas pressure in order to minimize the back reaction of cosmic rays on the gas. The simulation is performed with adaptive mesh refinement (AMR), with a minimum and maximum resolution of 1/256 and 1/2048, respectively, depending on the local second derivative of the CR energy density. Magnetic fields and CR diffusion are not included in this run.  

\begin{figure}[tp]
\begin{center}
\includegraphics[scale=0.7]{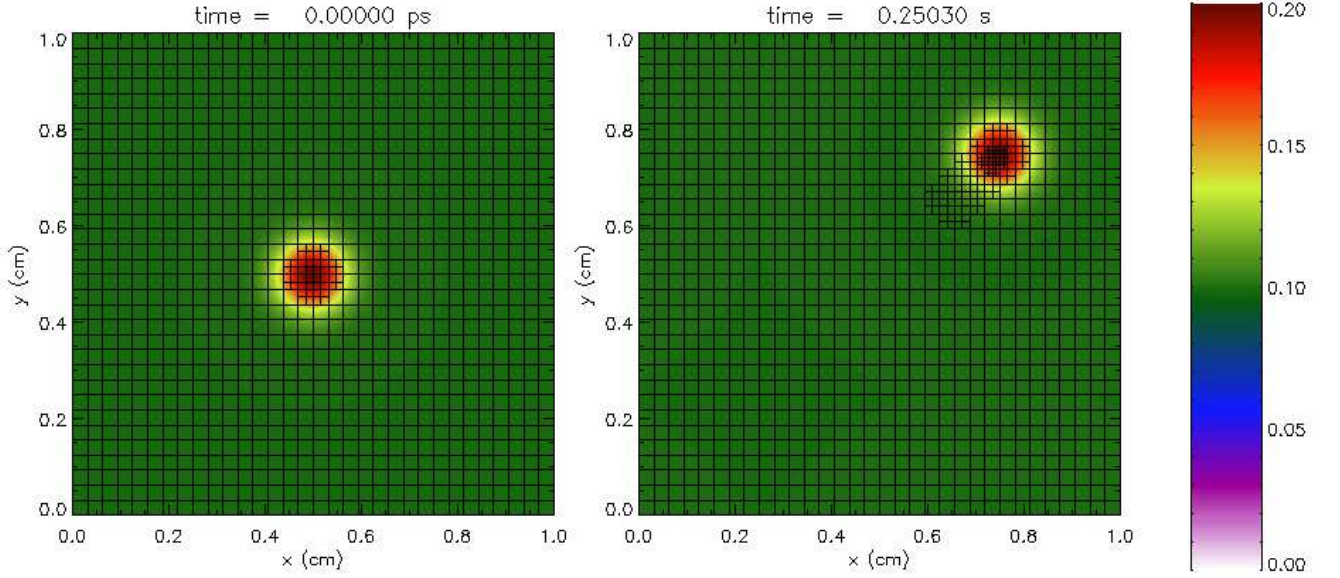} 
\caption{Test for CR advection. Left and right panels are the CR energy density at $t=0$ and $t=0.25$\ s. Adaptive mesh tracks the overdensity of cosmic rays without introducing spurious features.} 
\label{fig:adv}
\end{center}
\end{figure}

Figure \ref{fig:adv} shows the CR energy density at $t=0$ (left) and $t=0.25$\ s (right), overplotted with the locations of the AMR blocks (each block contains $8\times8$ grid cells). The cosmic rays passively evolve with the thermal gas along the diagonal direction without any distortion. Moreover, this test demonstrates that the AMR correctly traces the CR overdensities and does not introduce any artifacts.    

\subsection{A.2\ Tests for Magnetic Field Aligned CR Diffusion}

In order to verify the component of anisotropic CR diffusion in the new CR module, we put an overdensity of cosmic rays (Eq.\ \ref{eq:gauecr}) in two different magnetic field configurations: (1) diagonal field, and (2) loop field. In order to focus on the effect of CR diffusion, in these tests we turn off the calculation of hydrodynamic fluxes and only solve for the CR evolution due to diffusion along the field lines according to Eq.\ \ref{eq:anisotropic}. CR back reaction on the magnetic field is also neglected.

\begin{figure}[tp]
\begin{center}
\includegraphics[scale=0.7]{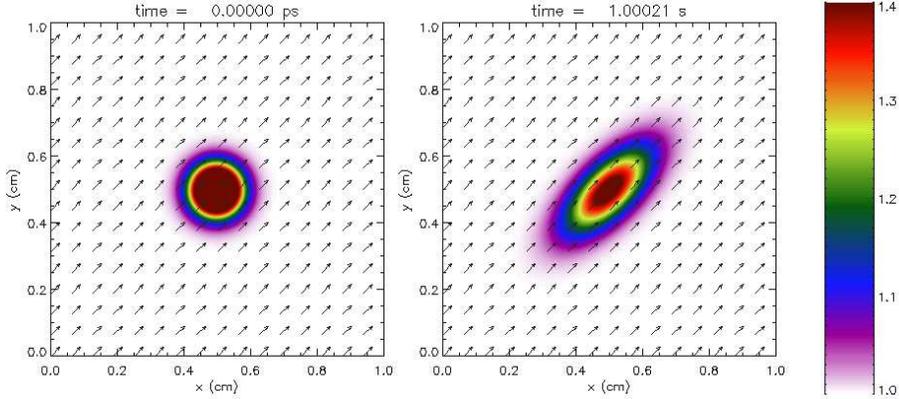} 
\caption{Evolution of the CR energy density due to diffusion along a diagonal magnetic field (arrows).} 
\label{fig:diagdiff}
\end{center}
\end{figure}

For the first `diagonal' test, the initial CR distribution is the same as in the advection test (see \S~\ref{sec:adv}), except that the fluids are at rest in the beginning of the simulation, and the cosmic rays are allowed to diffuse along the magnetic field with diffusion coefficients $\kappa_\parallel = 0.05$, and $\kappa_\perp = 0$. Here we present results for a simulation box of $1\times 1$ with a uniform grid of $256\times 256$ cells. Figure \ref{fig:diagdiff} displays the evolution of CR energy density in a diagonal magnetic field (arrows). The cosmic rays diffuse in the diagonal direction along the field lines as expected. 

\begin{figure}[tp]
\begin{center}
\includegraphics[scale=0.75]{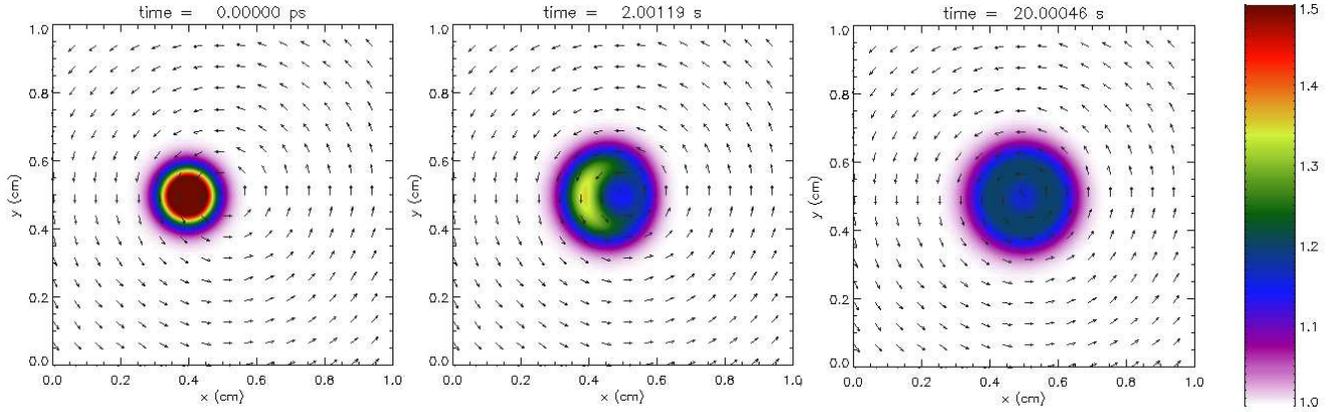} 
\caption{Time series of the distribution of the CR energy density. The cosmic rays gradually diffuse along the looped magnetic field (arrows).} 
\label{fig:loopdiff}
\end{center}
\end{figure}

Next, we show the test results of anisotropic CR diffusion along a `looped' magnetic field, which is more stringent because the field lines are inclined at all possible angles with respect to the Cartesian grid. For this test, we initialize a magnetic field with uniform amplitude $B_0=10^{-4} \mu$G in concentric circles around the center of the simulation box. This test is performed in 2D, and a virtual vector potential (${\bm B}=\nabla \times {\bm A}$) is used to ensure $\nabla \cdot {\bm B} = 0$. To generate a uniform looped field, the vector potential is chosen to be ${\bm A}=(A_x, A_y, A_z) = (0, 0, B_0(5-r))$, where $r$ is the radius from the center of the domain. The initial CR distribution is the same as in the diagonal test, but displaced by 0.1 from the center of the simulation box. The CR distribution at three different epochs is shown in Figure \ref{fig:loopdiff}. Due to anisotropic diffusion around the looped field (arrows), the cosmic rays start from a initially localized distribution and gradually diffuse following the field lines, and eventually equilibrate when the CR energy gradient in the azimuthal direction vanishes. This stringent loop test demonstrates the robustness of our implementation of CR anisotropic diffusion.

\subsection{A.3\ Linear Sound Wave Test for a Composite of Thermal Gas and Cosmic Rays}

\begin{figure}[tp]
\begin{center}
\includegraphics[scale=0.5]{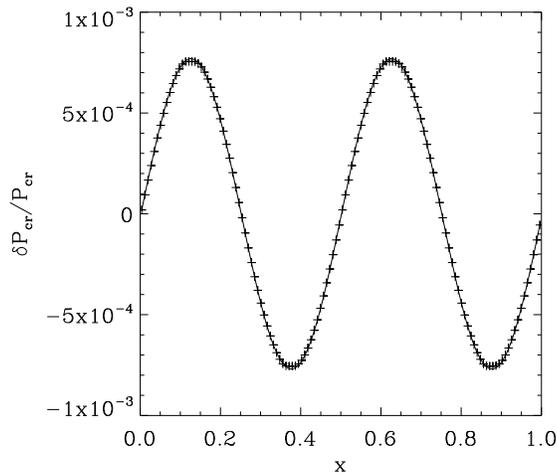} 
\caption{Linear sound wave test for a composite of thermal gas and cosmic rays. The figure shows the perturbation of the CR pressure along the $x$-direction at $y=0.5$ after one wave crossing time. Simulated data is plotted with the plus symbols; solid line shows the analytical solution.} 
\label{fig:soundwave}
\end{center}
\end{figure}

Next, we perform additional quantitative tests of the module. Here we present results of a 2D test of a sound wave propagating in a composite of thermal gas and cosmic rays \citep{Rasera08}. This test problem takes into account the coupling between the two fluids. The medium is perturbed by small fluctuations that satisfy the following relations 
\begin{eqnarray}
\frac{\delta \rho}{\rho_0} &=& \frac{\delta v}{c_s}, \\ \label{eq:drho}
\frac{\delta p}{p_0} &=& \gamma \frac{\delta v}{c_s}, \\
\frac{\delta p_{\rm cr}}{p_{\rm cr,0}} &=& \gamma_{\rm cr} \frac{\delta v}{c_s}, \label{eq:dpcr}
\end{eqnarray} 
where $\delta \rho$, $\delta v$, $\delta p$, and $\delta p_{\rm cr}$ and perturbations, $\rho_0$, $p_0$, $p_{\rm cr,0}$ are the initial unperturbed quantities, and the adiabatic wave speed is 
\begin{equation}
c_s = \sqrt{\frac{\gamma p_0+\gamma_{\rm cr}p_{\rm cr,0}}{\rho_0}}.
\end{equation}
For our test, we initialize a sine wave moving in the $x$-direction with wavelength 0.5 and amplitude $\delta v=10^{-3}$ in a medium with unperturbed quantities $\rho_0=1$, $v_0=0$, $p_0=1$, and $p_{\rm cr,0}=1$. The other perturbed variables are calculated according to Eq.\ \ref{eq:drho}-\ref{eq:dpcr}. All variables have values independent of their $y$ coordinates. The wave travels periodically in a simulation box of sizes $1\times 1$ with 128 grid cells on a side. Figure \ref{fig:soundwave} shows the results after propagating the sound wave for one period. Our results (plus signs) show excellent agreement with the analytical solution (solid line).    

\subsection{A.4\ Nonlinear Shock Tube Test for a Composite of Thermal Gas and Cosmic Rays}
The Sod shock tube problem \citep{Sod78} is a standard test for the accuracy of computational fluid codes, in particular the Riemann solver. The original problem contains a polytropic gas separated into a high pressure and high density left state and a low pressure and low density state on the right side, which then evolves into a system characterized by a rarefaction wave, a contact discontinuity, and a shock. The analytical solution for the propagation of these characteristics can be derived using the Rankine-Hugoniot jump conditions. 

Similarly, for the case of a hybrid fluid consisting of thermal gas and cosmic rays, an analytical solution of the shock tube problem has been derived by \cite{Pfrommer06}. We perform this test in 2D in a simulation box of size $1\times 1$ on a uniform grid with 1024 cells on a side. The variables of the left state ($0<x<0.5$) are initialized to be $\rho_L=1$, $v_L=0$, $p_L=6.7\times10^4$, and $p_{{\rm cr},L}=1.3\times 10^5$; the right state ($0.5<x<1$) has values of $\rho_R=0.2$, $v_R=0$, $p_R=2.4\times 10^2$, and $p_{{\rm cr},R}=2.4\times 10^2$. The values on the cells are independent of their $y$ coordinates. 

\begin{figure}[tp]
\begin{center}
\includegraphics[scale=0.8]{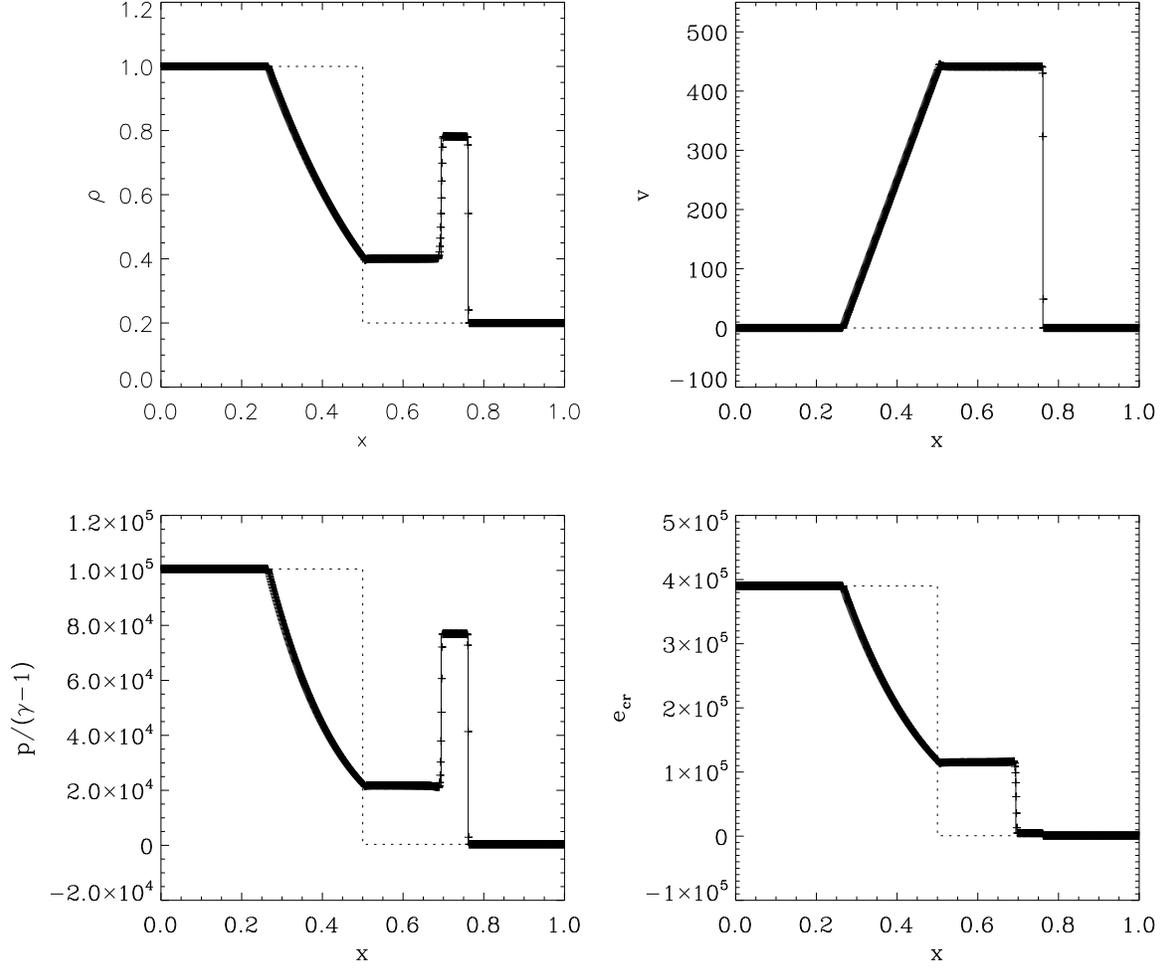} 
\caption{Shock tube test for a hybrid fluid of thermal gas and cosmic rays. The figure shows profiles of the density (top left), velocity (top right), gas internal energy density (bottom left), and CR energy density (bottom right) sliced through $y=0.5$. Plus signs are the simulated data, solid lines are the analytical solution, and dotted lines show the initial condition.} 
\label{fig:shocktube}
\end{center}
\end{figure}

The results after running the simulation for $t=4.4\times 10^{-4}$ are plotted in Figure \ref{fig:shocktube}. Again, the simulated results (plus signs) match very well with the analytical prediction (solid line). The transitions between characteristics are well located. The profile of the rarefaction wave is reproduced, and the contact discontinuity and the shock are resolved within few cells without spurious oscillations. Based on the above test results, we conclude that our implementation of the CR module in FLASH is successful.   

\end{document}